\documentclass[twocolumn,tighten]{aastex61}
\input psfig.tex

\usepackage{aas_macros}
\usepackage{latexsym}
\usepackage{natbib}
\usepackage{amssymb}
\usepackage{amsmath}

\usepackage{graphicx}
\usepackage{graphics}
\usepackage{fancyhdr}
\usepackage{morefloats}

\def\fp1{\mbox{$a\log(R_e)+b\log(<I_e>)+c\log(\sigma)+d=0$}}

\def\H0{\mbox{$H_0$}}
\def\q0{\mbox{$q_0$}}

\def\a4{$a_4/a$}

\def\D4000{$\Delta 4000$}

\def\M12{$10^{12}m_{\odot}$}

\def\lesssim{\,\lower2truept\hbox{${<\atop\hbox{\raise4truept\hbox{$\sim$}}}$}\,}
\def\gtrsim{\,\lower2truept\hbox{${>\atop\hbox{\raise4truept\hbox{$\sim$}}}$}\,}

\newcommand{\bq}{\begin{equation}}
\newcommand{\eq}{\end{equation}}

\def\p1{\partial}

\def\D{D_{\mbox{\em dif}}}

\def\smallskip{\vskip 8pt}

\shorttitle{Cosmic Star Formation: A Simple Model of the SFRD(z)}
\shortauthors{C. Chiosi et al.}

\begin{document}

\title[Cosmic Star Formation: A Simple Model of the SFRD(z) ]{Cosmic Star Formation: A Simple Model 
of the SFRD(z) }

\correspondingauthor{Cesare Chiosi}
\email{cesare.chiosi@unipd.it}

\author{Cesare Chiosi}
\affiliation{Department of Physics and Astronomy, University 
of Padova, Vicolo Osservatorio 3, I-35122 Padova, Italy}
\affiliation{INAF Observatory of Padova, Vicolo Osservatorio 5, I-35122 Padova, Italy}

\author{Mauro Sciarratta}
\affiliation{Department of Physics and Astronomy, University 
of Padova, Vicolo Osservatorio 3, I-35122 Padova, Italy}

\author{Mauro D'Onofrio}
\affiliation{Department of Physics and Astronomy, University 
of Padova, Vicolo Osservatorio 3, I-35122 Padova, Italy}
\affiliation{INAF Observatory of Padova, Vicolo Osservatorio 5, I-35122 Padova, Italy}

\author{Emanuela Chiosi}
\affiliation{INAF Observatory of Padova, Vicolo Osservatorio 5, I-35122 Padova, Italy}

\author{Francesca Brotto}
\affiliation{Department of Physics and Astronomy, University 
of Padova, Vicolo Osservatorio 3, I-35122 Padova, Italy}

\author{Rosaria De Michele}
\affiliation{Department of Physics and Astronomy, University 
of Padova, Vicolo Osservatorio 3, I-35122 Padova, Italy}

\author{Valeria Politino}
\affiliation{Department of Physics and Astronomy, University 
of Padova, Vicolo Osservatorio 3, I-35122 Padova, Italy}



\received{June 7, 2017}
\revised{}
\accepted{}

\submitjournal{ApJ}

\label{firstpage}

\begin{abstract}
We investigate the evolution of the cosmic star formation rate density (SFRD) from redshift z=20 to z=0 and
compare it  with the observational one by Madau and Dickinson derived from recent compilations of UV and
IR  data. The theoretical SFRD(z) and its evolution are obtained using a simple  model which
folds together the
star formation histories of prototype galaxies designed to represent real objects of different
morphological type
along the Hubble sequence and the hierarchical growing of structures under the action of gravity from small
perturbations to large scale objects in $\Lambda$-CDM cosmogony, i.e. the number density of dark matter 
halos $N(M,z)$.
Although the overall model is very simple
and easy to set up, it provides results that well mimic those obtained from large scale N-body simulations
of great  complexity. The simplicity of our approach allows us to test different assumptions for the star
formation law in galaxies, the effects of energy feedback  from stars to  interstellar gas and the
efficiency of galactic winds, and
also the effect of $N(M,z)$. The result of
our analysis is that in the framework of the hierarchical assembly of galaxies 
the so-called time-delayed star
formation  under plain assumptions mainly for the  energy feedback and galactic winds   
can reproduce  the observational SFRD(z).
\end{abstract}



\section{Introduction}\label{Intro}

This paper faces the problem of the global history of star formation  and 
chemical enrichment of the whole Universe,
otherwise known as the baryon budget in galactic halos or the history of the 
so-called star formation rate density SFRD(z). Since the seminal studies by \citet{Tinsley1980} and 
\citet{Madau_1996}, the cosmic star formation has been the subject of numberless papers that are impossible
to recall here.   
The evolution of the SFRD(z) over cosmic times is crucial to understand galaxy formation
and evolution and to constrain any theory devoted to this subject \citep{Hopkins2004, Wilkinsetal2008,
Guoetal2011, Bouwensetal2012, Cucciatietal2012, Tescarietal2014, KatsianisTescarietal2017, Abramsonetal2016}.
The evolution of the SFRD(z) is nowadays known with unprecedented accuracy up to the distant
Universe thanks to the multi-wavelengths surveys
carried out by many groups among which we recall
\citet{Bernardietal2010, Gonzalezetal2011, Bouwensetal2012, Lee_etal2011,Smitetal2012, Santinietal2012,
Schenkeretal2013,vandenBurgetal2010,Gruppionietal2013,Parsaetal2016,Reddyetal2008,
Magnellietal2011,Sobraletal2013,Alavietal2014,Cucciatietal2012,Lyetal2011}.
The situation has been recently systematically summarized and reviewed by \citet{MadauDickinson2014} and
\citet{KatsianisTescarietal2017} to whom we refer for all details.

The interpretation of  the cosmic SFRD(z) has been addressed by many theoretical studies, among
which we recall
\citet{RaseraTeyssier2006}, \citet{HernquistSpringel2003}, and \citet{KatsianisTescarietal2017}, with either
analytical or  semi-analytical or hydrodynamical simulations. In particular, they investigated the effect 
of the energy feedback from supernovae explosions,
stellar winds, and AGN activity on modeling the cosmic star formation. They made use of an improved version
of the P-GADGET3 of
 \citet{Springel2005} with chemical enrichment \citep{Tornatore-etal2007}, supernova energy and
 momentum-driven galactic winds
\citep{PuchweinSpringel2013}, AGN feedback \citep{SpringelDiMatteo2005,Planellesetal2013}, metal-line cooling
\citep{Wiersma2009a,Wiersma2009b} plus
molecules/metal cooling \citep{Maioetal2007}, supernova-driven galactic winds with feedback
\citep{Baraietal2013}, thermal conduction \citep{Dolagetal2004},  and other more technical details
\citep[see][for a more exhaustive description]{Tescarietal2014}. In general the shape of the SFRD(z)
as a function of the redshift is reproduced by the models. However, according to \citet{Tescarietal2014} the
SFRD(z) is insensitive to feedback at $z > 5$, unlike to what happens at lower redshifts.
They find that the key factor
for reproducing the observational SFRD is a combination of strong supernova-driven wind and early AGN
feedback
in low mass galaxies. Similar conclusions are reached by \citet{KatsianisTescarietal2017} in the sense
that the AGN feedback
they adopt decreases the SFRD at $z <3$  but not sufficiently at  higher redshift.  According to them,
the kind of feedback one would need to reconcile things is a strong feedback at high redshifts and
a less efficient one at low redshift. They also  show that variable galactic winds, which are efficient
at decreasing the star formation rate (SFR) of low mass galaxies, are  quite successful in reproducing
the observational data.

\textsf{Aims of this study}. Although the theoretical SFRD(z) obtained in those studies nicely 
reproduces the observational one (which is not a
surprise since some important physical ingredients,  such as for instance the energy feedback from
AGNs via the
galactic winds on a galaxy's SFR, have  yielded the sought variation of the
cosmic SFRD with the redshift), still we feel that the results are not yet conclusive as far as the key
physical process in shaping the cosmic SFRD(z) is concerned. Casting the question in a different way,
we would like  to understand whether the cosmic SFRD(z) is driven more by causes of external or internal
nature. 

Among the external causes, chief is the gravitational building up of structures (the proto-galaxies
made of dark  and baryonic matter) via hierarchical aggregation which leads to a mass function of galaxies
which is not the same at different redshifts.
The numerical
simulations  of cosmic mass aggregation show that the halo mass distribution function, i.e. the relative number of
galaxies per mass interval, on one hand gets steeper and steeper with mass at increasing redshift but even 
more important several different solution are found \citep[and references therein]{Murray_etal2013} all worth 
being explored.

Among the internal causes, chief are the star formation,
how this varies with the total mass and the mean density of the galaxy, how the SFR varies
with time within  a galaxy, and the physical properties of the interstellar medium. Another important issue
is whether in a galaxy  the SFR always starts at maximum efficiency and declines with time so that some
``quenching mechanisms'' must be invoked at the very early epochs to explain the decline of the 
SFRD(z) at increasing redshift, or rather
it starts low, grows to a maximum and then declines (typical of spheroidal systems), or alternatively it 
remains mild and nearly constant (such as in disk-like objects),
or in some cases it goes
through a series of episodes (the so-called bursting model, typical of low mass galaxies). 

Finally, we would
like to quantify the relative weight of the hierarchical aggregation compared to the intrinsic SFR. Most
likely both  concur to model the SFRD(z),  but to which extent?   Investigations based  on large scale
numerical simulations that are possible with P-GADGET3-like codes have the gravitational aggregation
built in by default so that only the effect of different prescriptions for the star formation and
associated energy injection and feedback can be tested. On the other hand, testing the effects of the various 
physical ingredients with direct hydrodynamical large scale simulations is expensive  and time consuming. 
For these reasons it is still useful and interesting to address the problem in a simple fashion by means of 
semi-analytical model able to catch the essence of the problem.   

The plan of the paper is as follows.
In Section \ref{CSFR} we shortly review the present-day observational picture about the SFRD(z),
recalling the main ways of measuring it, the various points of uncertainty, in particular the role
of the stellar initial mass function (IMF), the variation of the star formation histories 
in galaxies of different type, and the distance determinations.  
In Section \ref{strategy} we present the strategy of the
present study aimed at deriving
the SFRD(z) from three elementary building blocks: (i) the current hierarchical view of
galaxy formation providing the expected number of galaxies (made of dark and baryonic matter)
per unit volume (usually a $Mpc^3$) presented in Section \ref{first_block}, (ii) simple models of galaxy
formation and  evolution  for different values of total mass and morphological type that  are
presented in Section \ref{second_block} (they provide the rate
of star formation, mass in stars, gas content, metallicity and other useful properties of
 individual galaxies),  and  (iii) finally, the evolutionary population synthesis
technique that is used to derive the  magnitudes and colors of the model galaxies as function of time
(redshift) that is presented in Section \ref{third_block}. 
In Section \ref{SFRD_theory}
we fold together the results of the previous sections,  derive the cosmic SFRD and compare it to
observational data. In order to highlight and single out the role played by the galaxy number
density distribution and the galaxy SFR at different epochs, we perform some ad hoc
simulations by varying some key assumptions and illustrate the results.
Finally in Section \ref{Conclusions} some conclusive remarks are made.

\begin{figure*}
\centering{
 {\includegraphics[width=7.5cm]{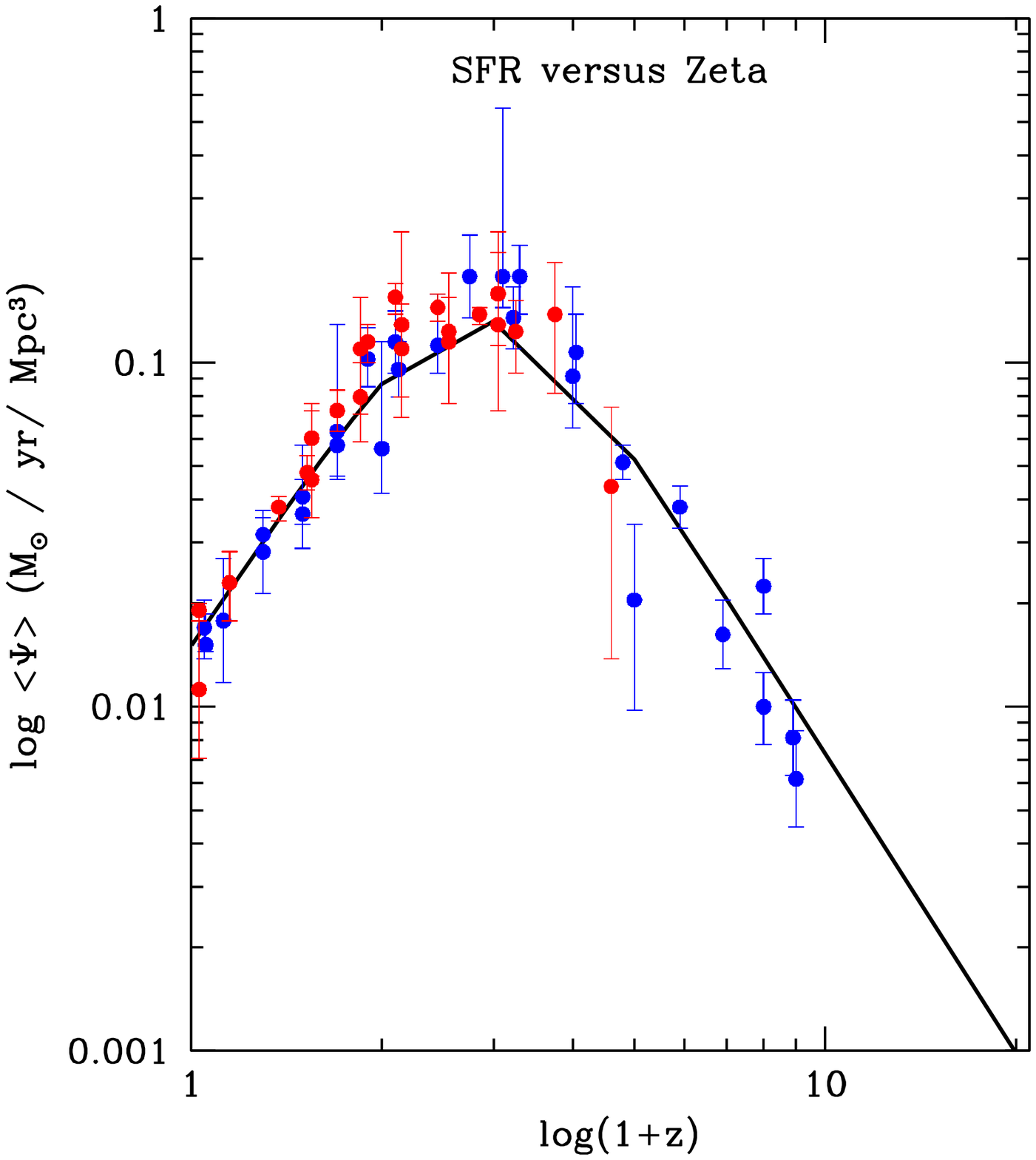}  }
 {\includegraphics[width=7.5cm]{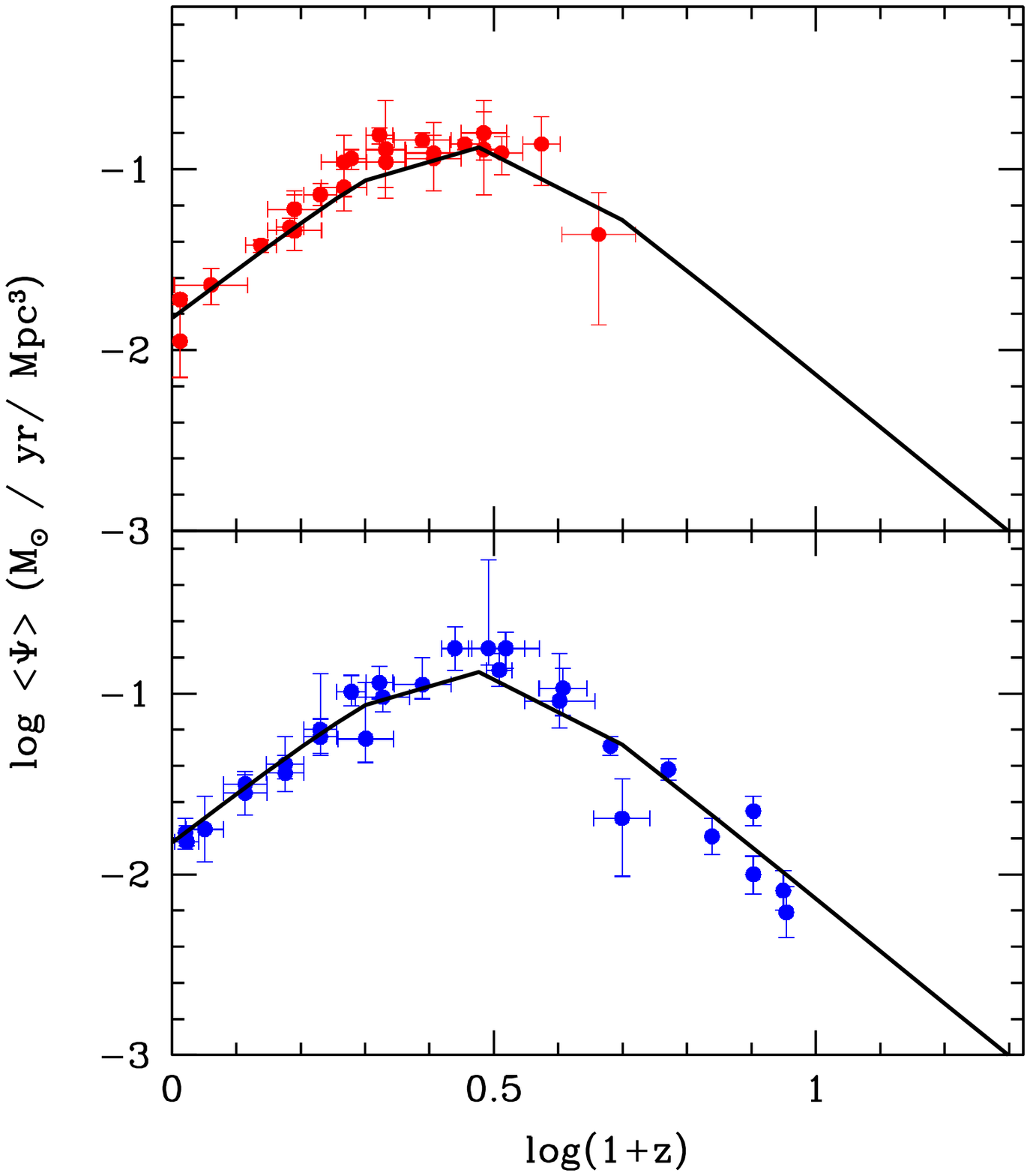} }   }
\caption{ The history of cosmic star formation according to
\citet[][their Fig.8]{MadauDickinson2014}. The left panel shows the rest-frame FUV+IR data 
(blue and red dots respectively), whereas in the right panels the same data are plotted separately. The 
sources of data are those listed  in Table 1 of
\citet{MadauDickinson2014}.  
The solid line in the
three panels is the analytical best fit of the data given by \citet{MadauDickinson2014}.
}
\label{fig_madau_dickinson}
\end{figure*}

\section{The cosmic star formation rate}\label{CSFR}

The SFRD(z) we intend to investigate and reproduce is the one presented by 
\citet[][and references therein]{MadauDickinson2014}. In general, to infer the SFRD(z) from the fluxes measured in suitable pass-bands (typically UV and 
near and far infrared, NIR and FIR, respectively) and to express it  
in masses per unit time and unit volume of space, one needs some   assumptions about the correlation 
between the measured fluxes and the SFR, the corrections for the effect of dust on absorbing
part of the UV to re-emit it in the NIR and FIR, the IMF together 
with some hints about its constancy or variation with time and space,  the kind of star formation at work 
on cosmic scales and Hubble time, and others details.

\textsf{Major uncertainties}. A number of problems affect the determination of the SFRD(z) among 
which we briefly recall:

\textit{Stellar mass census}. Deriving the mass in stars (i.e. the underlying IMF)
from their light is a cumbersome affair
because it requires information on the  mass to light ratio  (M/L) of the
stellar populations, which in turn depends on the age, the history of star formation and the amount of
dust around (extinction). 
In general, the conversion from light to mass is made through population synthesis models
which provide  the relationship between mass in stars 
 (both luminous and faint - hence invisible), the light
emitted by these, and the relative number of stars born in different generations,   all
contributing to the light and the mass at present day; in other words, the history of star
formation.  Among other things, these models depend  on the IMF.  On the other hand, 
the IMF is difficult to determine directly from the observational data for a number of reasons that do not
need to be examined here \citep[see][for a detailed discussion of the issue]{MadauDickinson2014}. 
The obvious way out is to assume a certain IMF. The most popular one is the 
\citet{Salpeter1955} law
even though it is known to predict M/L ratios higher
than observed, thus requiring deep revision of the IMF at the low mass hand
\citep[see][]{Kroupa_etal_1993,Chabrier2015, HennebelleChabrier2011}.
Another difficulty affecting the stellar mass census is due to the detection of low mass dusty
galaxies. This means that  a great portion of the stellar mass could be missing in current data.

\textit{Variations of the star formation history in galaxies}. The star formation history (SFH) of 
a galaxy  may change a lot over the years
both on short and long timescales. As  a matter of fact, young stars outshine the old ones thus
affecting the total spectral energy distribution, and hence the total mass of the old stars may be largely
underestimated and the effect of these can be hardly singled out.

\textit{Distances}. Finally, the sources of observational data change with the distance so that
homogeneous data sets extending from the local pool all the way up to redshift $z \leq 10$ are not possible:
For instance in the local Universe  ($0<z<1$) most of the IR data are not due to dust in star-forming
regions but to dust in the ISM.  This trend  tends to decrease with the distance. In the redshift interval
$1<z<4$, no IR data are measured for individual sources but for the hyper-luminous ones, 
thus heavily affecting 
 the evaluation of  the  IR luminosity density. At larger redshifts essentially only data for 
 hyper-luminous sources are available thus worsening the problem. 
For all these reasons, \citet{MadauDickinson2014}  limit their analysis to the redshift interval
$0<z<8$.

\textsf{Analytical Fits}. In this work, we will make use of the analytical fits derived by 
\citet{MadauDickinson2014} and \citet{MadauFragos2017}. Both have  similar functional dependencies 
 given by
		
\begin{equation}
SFRD(z)=\gamma_0 \frac{(1+z)^{\gamma_1}}{ 1 + (\frac{1+z}{\gamma_2})^{\gamma_3} }\,\,
M_\odot yr^{-1} Mpc^{-3}
\label{eq_madau_dickinson}
\end{equation}
The relation of \citet{MadauDickinson2014} is for $\gamma_0=0.015$, $\gamma_1=2.7$, 
$\gamma_2=2.9$, and $\gamma_3=5.6$ and it is shown  in the three panels of
Fig.\ref{fig_madau_dickinson} together with the original data from the same source. 

In recent times, the above relationship has been slightly revised by
\citet{MadauFragos2017} who used the \citet{Kroupa2001} IMF. 
The new coefficients and exponents are 
$\gamma_0=0.01$, $\gamma_1=2.6$, $\gamma_2=3.2$, and $\gamma_3=6.2$. It is easy to check 
that the old and new relationships agree within a factor of about two. 
Both represent the foot-print of the past star and galaxy formation history of the Universe
that needs to be deciphered (see Fig.\ref{two_fits} for a comparison).

\section{Strategy:  deriving the SFRD(z) from fundamental building blocks}\label{strategy}

In this study, we intend to derive the observational SFRD(z) from a small number of hypotheses
or ``building blocks'': 

1) The cosmic scenario and the hierarchical building up of bound structures
which provide the number density of DM halos of  mass $M_{DM}$ and  radius
$R_{DM}$ as function of the redshift, $N(M_{DM,} z)$.

2) The aggregation of BM  in  DM halos which provides the visible component of galaxies and their star 
formation and chemical enrichment. This gives rise to a complicate  game among several important 
physical processes, chief among others the gravitational  contraction and collapse together 
with gas heating and  cooling and star formation. All this requires a suitable timescale 
to occur so that building up of the stellar component of a galaxy  cannot be
instantaneous. The best simple
model apt to describe this situation is the so-called ``infall model'' developed by \citet{Chiosi1980}.

3) The spectro-photometric properties of the stellar population of galaxies that will provide the 
evolution of spectral energy distribution as function of time, SFH and 
chemical enrichment.
This gives us  magnitudes and colors of the stellar populations in galaxies as a
function of the time and/or redshift
for whatsoever photometric system in use.

The SFRD(z) results  from folding together the building blocks above: at each redshift we know
the number density of DM halos and associated BM
galaxies born in the redshift  $z_{f} \leq z$, where $z_{f}$ is the  redshift at which the
first galaxies
are supposed to form ($z=20$ in our case).  At each redshift we calculate the number density
of galaxies per $Mpc^3$ as a function of the mass of the DM halo (this soon sets the mass of the
 BM galaxy hosted by
a DM halo). For this ideal sample of
galaxies we calculate the total and mean  cosmic density of star formation, metallicity, mass in stars,
and luminosity emitted in any 
pass-band according to

 \begin{equation}
           [ \mathcal{F} ]_{T}   =
         \int \int \mathcal{F}(M_{DM}, z, z_{f} )
           \times N(M_{DM},z,z_{f} )dM_{DM} dz     \\
\end{equation}

\begin{equation}
< \mathcal{F} > = \frac
        { \int \int \mathcal{F}(M_{DM}, z, z_{f})
           \times N(M_{DM},z,z_{f})dM_{DM} dz }
                  {\int \int N(M_{DM}, z, z_{f}) dM_{DM}  dz}
 \end{equation}

\noindent
where $\mathcal{F}$ stands for any of the physical quantities listed above and
the integrals are carried out over the range of $M_{DM}$ and  $z_{f} \geq z \geq 0$ we have considered.
The correspondence between the halo mass $M_{DM}$ and the BM galaxy mass $M_{BM}$
inside is fixed by  the cosmological model of the Universe (see below).

\begin{table*}
\begin{center}
\caption{ Coefficients of the polynomial interpolation of the relation (\ref{Lukic_interp}),
which provides the number density of halos $N(M_{DM}, z)$ per (Mpc/h)$^3$.}
\label{coef_lukic}
\begin{tabular}{|c|r|r|r|r|r|}
\hline
Mass $[M_\odot/h]$  &     A$_4$      &    A$_3$       &    A$_2$      &    A$_1$       &    A$_0$ \\
\hline
 5e7                &-2.34275e-5     &   1.28686e-3   & -2.97961e-2   &   2.11295e-1   &  2.02908  \\
 5e8                &-2.76999e-5     & 	 1.49291e-3   & -3.47013e-2   &   2.13274e-1   &  1.13553  \\
 5e9                &-1.31118e-5     & 	 6.50876e-4   & -2.36972e-2   &   1.31993e-1   &  0.23807  \\
 5e10               &-1.18729e-5     & 	 6.65488e-4   & -3.17079e-2   &   1.30360e-1   & -0.59744  \\
 5e11               &-1.47246e-5     & 	 8.10097e-4   & -4.65279e-2   &   1.13790e-1   & -1.44571  \\
 5e12               & 6.59657e-5     &  -7.19134e-4   & -6.99445e-2   &   1.06782e-1   & -2.45684  \\
 5e13               &-7.34568e-4     &   9.99022e-3   & -1.65888e-1   &  -9.48292e-2   & -3.11701  \\
\hline
\end{tabular}
\end{center}
\end{table*}

\section{First building block:  number of DM halos  at different
redshifts}\label{first_block}

We assume the $\Lambda$-CDM concordance cosmology, with values inferred from the
WMAP-5 data \citep{Hinshawaetal2009}:
flat geometry, $H_0=70.5$ km/s/Mpc, $\Omega_{\Lambda} = 0.72$, $\Omega_m=0.28$,
$\Omega_b=0.046$
(giving a baryon ratio of $ \Omega_b/\Omega_m \simeq 0.1656$), $\sigma_8=0.817$, and $n=0.96$.
To these values
for $\Omega_m$ and $\Omega_b$ we have the corresponding ratio between the baryonic and dark matter
masses of individual galaxies
$M_{BM} /M_{DM} \simeq 0.16$ and vice versa  $ M_{DM} / M_{BM} \simeq 6.12$.

As already mentioned, the standard approach to investigate the cosmic SFRD is based on large scale
cosmological N-Body simulations
 in the framework of a given cosmological model of the Universe ($\Lambda$-CDM in our case)
so that the appearance, growth
and subsequent aggregation of perturbations of all scales can be suitably described
\citep[e.g.]{Springeletal2005}. The formation of DM halos and BM galaxies inside are automatically taken
into account in the simulations. The price to pay is a large computational cost so that
the analysis is limited to a few paradigmatic cases.

In alternative, one may adopt the strategy used by \citet{Lukicetal2007}. 
Starting from  the \citet[][]{Warren_etal_2006} 
halo mass function (HMF), they  derive the
halo growth function (HGF) in the concordance $\Lambda$-CDM model over a wide range of redshifts
(from $z\simeq 20$ to the present) (see their Fig.2). The HGF  $N(M_{DM}, z)$ 
gives the number density of halos of different masses  per (Mpc $h^{-1})^3$ resulting
by all creation/destruction events. 
By performing a large
suite of nested-box N-body simulations with careful convergence and error controls,
they determine the mass
function and its evolution with excellent statistical and systematic
errors, reaching a few percent over most of the considered redshift and mass range.
The advantage of the  
\citet{Lukicetal2007} study is that it provides
a halo mass distribution function, $N(M_{DM}, z)$, easy to use in all cases like the present one
in which galaxy evolution has to be framed in a cosmological context.

In order to make use of the \citet{Lukicetal2007}
distribution 
in our analysis, we fit their results  with a fourth order polynomial

\begin{equation}
\log N(M_{DM}, z) = \sum_{j=0}^4 A_j(M_{DM}) \times z^j \label{Lukic_interp}.
\end{equation}

\begin{figure}
\centering
\includegraphics[width=8.5cm]{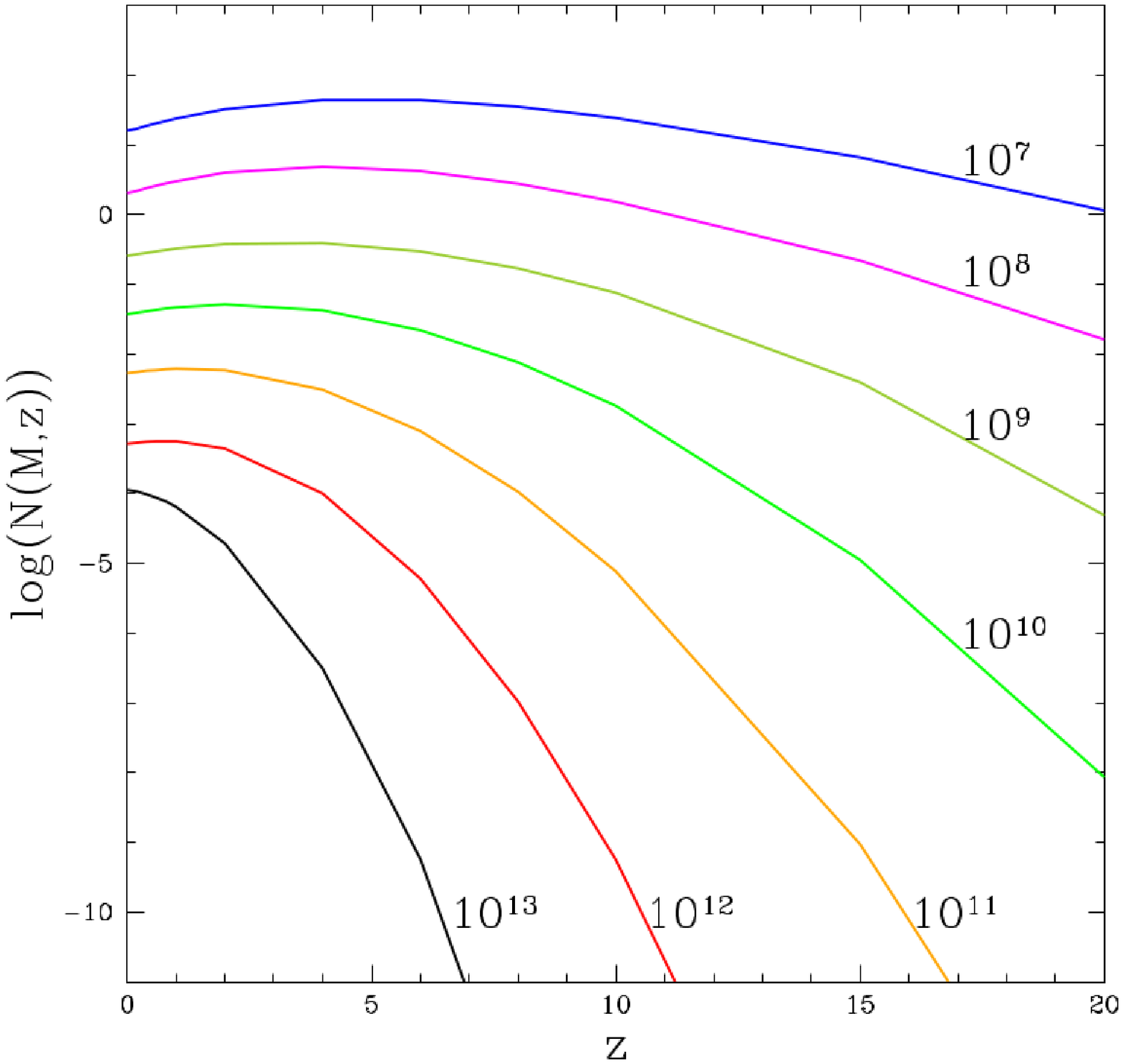}
\caption{ The HGF reproduced
from \citet{Lukicetal2007}. The number density of galaxies (in logarithmic scale) is in (Mpc $h^{-1})^3$, where
 $h=H_0/100$. Each line refers to halos with DM mass in solar units as indicated.}
\label{fig_lukic}
\end{figure}

\noindent The coefficients $A_j(M_{DM})$ are listed in Table \ref{coef_lukic}. 
The interpolated distribution function  for the
number density $N(M_{DM},z)$ of halos per $Mpc^3$  as a function of the mass and redshifts is shown here
 in Fig. \ref{fig_lukic}. As expected it is identical to the original one by \citet{Lukicetal2007}.

Although what we are going to say is well known, see the pioneer study of
\citet{PressSchechter1974} and \citet[][ for ample referencing]{Lukicetal2007}, for the sake
of clarity and relevance for our discussion we note the following: (i) for each halo mass
(or mass interval) the number density is small at high redshift, increases to high values towards the 
present,
 and depending on the halo mass either gets a maximum value at a certain redshift followed by a decrease
(typical of low-mass halos) or it keeps increasing as in the case of high-mass halos; in other words,
first creation of halos of a given mass (by spontaneous growth of perturbation to the collapse regime or
by mergers) overwhelms their destruction (by mergers), whereas the opposite occurs past a certain value
of the redshift, for low mass halos; (ii) at any redshift high mass halos are orders of magnitude less
frequent than the low mass ones; (iii) at any redshift, the mass distribution of halos has a typical
interval of existence whose upper mass end (cut-off mass) increases at decreasing redshift.

Finally it worth recalling that both the number densities $N(M_{DM},z)$ \citet{Lukicetal2007} and the 
SFRD(z) of   \citet{MadauDickinson2014} are per $Mpc^3$ so that
comparing theory with observations is less of a problem. However, owing to the many uncertainties
affecting the
observational data and the crudeness of the theoretical models, small adjustments of the order
of a few units  can be tolerated in the final comparison.

\begin{table*}
\begin{center}
\caption{ Expected number densities of DM halos per $ \, [Mpc\, h^{-1}]^{3}$
at varying the DM mass and redshift $z$. The DM masses are  in $M_{\odot}h^{-1}$.}
\label{count_lukic}
\begin{tabular}{|c c c c  c c c   c| }
\hline
  z          &$10^7$         &  $10^8$    &   $10^9$  &  $10^{10}$  &$10^{11}$&  $10^{12}$  &$10^{13}$  \\
  \hline
      20.00  &   0.113E+01 &  0.159E-01 &  0.482E-04 & 0.857E-08 &  0.346E-14  & 0.506E-24  & 0.000E+00 \\
      15.00  &   0.658E+01 &  0.215E+00 &  0.390E-02 & 0.111E-04 &  0.930E-09  & 0.335E-16  & 0.120E-45 \\
      10.00  &   0.240E+02 &  0.152E+01 &  0.753E-01 & 0.180E-02 &  0.762E-05  & 0.566E-09  & 0.176E-18 \\
        8.00 &   0.347E+02 &  0.274E+01 &  0.168E+00 & 0.756E-02&   0.104E-03 &  0.103E-06 &  0.691E-13 \\
        6.00 &   0.433E+02 &  0.415E+01 &  0.293E+00 & 0.219E-01&   0.778E-03 &  0.601E-05 &  0.568E-09 \\
        4.00 &   0.433E+02 &  0.485E+01 &  0.389E+00 & 0.419E-01&   0.304E-02 &  0.994E-04 &  0.312E-06 \\
        2.00 &   0.319E+02 &  0.395E+01 &  0.376E+00 & 0.507E-01&   0.585E-02 &  0.437E-03 &  0.190E-04 \\
        1.00 &   0.235E+02 &  0.299E+01 &  0.322E+00 & 0.461E-01&   0.610E-02 &  0.554E-03 &  0.638E-04 \\
        0.80 &   0.218E+02 &  0.278E+01 &  0.309E+00 & 0.445E-01&   0.601E-02 &  0.559E-03 &  0.755E-04 \\
        0.60 &   0.202E+02 &  0.257E+01 &  0.295E+00 & 0.427E-01&   0.586E-02 &  0.556E-03 &  0.869E-04 \\
        0.40 &   0.185E+02 &  0.237E+01 &  0.280E+00 & 0.408E-01&   0.568E-02 &  0.546E-03 &  0.973E-04 \\
        0.20 &   0.169E+02 &  0.217E+01 &  0.266E+00 & 0.387E-01&   0.545E-02 &  0.529E-03 &  0.106E-03 \\
        0.00 &   0.157E+02 &  0.201E+01 &  0.254E+00 & 0.371E-01&   0.525E-02 &  0.512E-03 &  0.111E-03 \\
\hline
\end{tabular}
\end{center}
\end{table*}

\section{Second building block: the BM  Galaxies inside DM halos}\label{second_block}

Given the mass distribution of DM halos as a function of the redshift and knowing the mass $M_{BM}$
of baryons inside thanks to the cosmological proportions $M_{BM}/M_{DM}  \simeq 0.16$, one
 needs a
prescription to get a model galaxy out of this lump of matter. DM and BM undergo gravitational collapse,
baryons cool down,   accumulate towards the core, and form stars. The visible galaxy is gradually  built up.
The timescale needed to get to the stage of nearly complete generation of  the stellar content goes
from the typical free-fall time (say about 0.5 Gyr or so) to significantly longer than  this
(say about 2 Gyr or even longer) depending on the galaxy type (mass).
The NB-TSPH simulations of \citet{ChiosiCarraro2002} in a monolithic-like scheme and those  of
\citet{MerlinChiosi2006,MerlinChiosi2007,Merlinetal2012} in the early hierarchical one
show that at decreasing galaxy mass the SFR shifts from a single prominent early episode to
ever continuous bursting-like
mode as the galaxy mass and/or the  over-density of the initial perturbation decreases, and that in
massive and intermediate mass galaxies (with $M_{BM}$ from $10^{10}$ to $10^{11} M_\odot$ or more)
the building-up of the stellar component is complete up to 90\% or so before $z \simeq 2$. These overall
trends and timescales of galaxy formation have been independently found and confirmed by
\citet{Thomasetal2005} from their analysis of the line absorptions indices in a  large sample of galaxies.
See also the review of \citet{Renzini2006}.

The infall models of galactic chemical evolution over the years have reached a
very high degree of complexity and sophistication,  have been applied to study galaxies of different
morphological type going from early types to disks and irregulars and have proved to successfully explain
 many observational properties of galaxies such as  chemical abundances, gas and stellar content and,
with the aid of  photometric synthesis tools, also magnitudes and colors. The situation has been widely reviewed by
\citet{Matteucci2012,Matteucci2016}: we limit ourselves
to mention here those developed by \citet{Bressanetal1994,Chiosietal1998,Tantaloetal1996,Tantaloetal1998}
for early type galaxies, and by
\citet{PortinariChiosi1999,PortinariChiosi2000,Fattore2009} for spherical and disk galaxies with radial
flows of gas. 

In the following, we will use models adapted
to the one zone description (fully adequate to our purposes) from those elaborated
by \citet{Tantaloetal1998} in spherical symmetry. Over the years many important physical phenomena 
have been incorporated
in the chemical models, for instance gas heating by supernova explosions (both type II and type Ia),
stellar winds, gas cooling by radiative emission, in order to correctly evaluate the thermal content
of the gas eventually triggering the galactic winds, and finally the radial flows of gas. Only recently,
the same physical processes  have been included also in the N-body simulations of galaxy formation
and evolution. Due to the scarce communication between the two scientific communities,  the strong
predicting power of the costless chemical models with respect to the heavy time consuming numerical
simulations has been ignored.

The essence of all infall models  is the assumption of the gas accretion into the central
region of the
proto-galaxy at a suitable rate (driven by the timescale $\tau$)  and of gas consumption by
a Schmidt-like law of star formation. The gas accretion and consumption coupled together give rise to a
time dependence of the SFR closely resembling the one resulting from the N-body
simulations and the line absorption indices diagnostics. We will come back to this important issue later on
in this study.

In the framework of infall models, the luminous mass $M_{BM}$
increases with time according to

\begin{equation}
  \frac{dM_{BM}(t)} { dt} = {\dot{M}}_{BM,0}\,exp(-t/\tau)
\label{inf}
\end{equation}
where $\tau$ is the accretion timescale.
The constant $ \dot{M}_{BM,0}  $
is obtained from imposing that at the galaxy age
${T_{G}}$ the value ${M_{BM}(T_{G})}$ is reached:

\begin{equation}
  \dot{M}_{BM,0}  = \frac{M_{BM}(T_{G})} {\tau [1 - exp(- T_{G}/\tau)]}
\label{mdot}
\end{equation}

\noindent
Therefore, integrating the accretion law the time dependence of
${M_{BM}(t)}$ is

\begin{equation}
 { M_{BM}(t) =  { \frac{M_{BM}(T_{G})}  {[1-exp (-T_{G}/\tau)] }    }
                      [1 - exp(-t/\tau)]  }
\label{mas-t}
\end{equation}
The above formalism allows us to immediately recover the {\it closed
box} approximation letting  $\tau \rightarrow 0$. The timescale $\tau$
parameterizes the timescale over which the present-day mass $M_{BM}(T_{G})$
is reached. In this scheme the total mass of a galaxy at the present time is
$M_G = M_{DM} + M_{BM}(T_G)$.

\subsection{ Basic equations of the model}

We denote with ${X_{i}(t)}$ the current mass abundance of an
element $i$ and introduce the dimensionless variables

\begin{equation}
{ G(t)=M_{g}(t)/M_{BM}(T_{G})  }
\label{gas_fra}
\end{equation}

\noindent
 and

\begin{equation}
{ G_{i}(t)=G(t)X_{i}(t),   }
\label{gas_i}
\end{equation}

\noindent
where by definition ${\Sigma_i X_i(t)=1}$.

The equations governing the time variation of the ${G_{i}(t)}$ and hence
${X_{i}(t)}$ are

\begin{eqnarray}\label{degas_i}
\frac{dG_{i}(t)}{dt} &=& - X_{i}(t) \psi(t) +    \nonumber \\
&&  [ {d G_{i}(t) \over dt} ]_{star}  +  [ {d G_{i}(t) \over dt} ]_{inf}
-  [{d G_{i}(t) \over dt }]_{win} 
\end{eqnarray}
where $\psi(t)$ is the normalized rate of star formation to be defined, and $t$  the current galaxy age. 

The four terms at the right hand side  are in the order: the rate of gas consumption by star formation, 
the rate 
of gas restitution (ejecta) by stars formed in previous epochs, the rate of mass 
accretion by infall of primordial gas 
onto the system, and the finally the rate at which enriched gas leaves the system.  
The infall rate is given by 

\begin{equation}
 [ { dG_{i}(t) \over dt } ]_{inf} = { X_{inf} \over M_{BM}(T_G) }  [ {dM(t) \over dt } ]_{inf}
\label{gas_inf}
\end{equation}
\noindent
and it is easily derived from eqn.(\ref{inf}). The rate of gas ejection is formally given by

\begin{equation}
 [ { dG_{i}(t) \over dt } ]_{win} = { X_{i}(t) \over M_{BM}(T_G) }  [ {dM(t) \over dt } ]_{win}
\label{gas_wind}
\end{equation}
is usually taken to be very high
(nearly instantaneous ejection of all heated up gas).

The rate of gas ejection by stars is  more complicated to calculate. The correct definition of this quantity
can be found in \citet{Bressanetal1994}, \citet{Tantaloetal1996}, and \citet{Tantaloetal1998}.
Suffice here to mention that (i) it requires integration over the 
IMF to 
account for the different contribution from stars of different mass and lifetime $\tau_M$, (ii) 
the stellar yields are 
calculated according to the so-called Q-formalism \citep[cf.][]{TalbotArnett1971,TalbotArnett1973}; (iii) 
at any age $t$, the rate of star $\psi(t)$ weighting the contribution  from star of different mass $M$ 
must be evaluated at $t_M=t - \tau_M$. The inclusion of Type Ia supernovae is made according to  
\citet{MatteucciGreggio1986} and it requires the specification of the mass interval and mass ratios 
for the binary stars progenitors of Type Ia supernovae together with the distribution function  
 $f(\mu)$  of their mass ratios and the percentage of such binary systems with respect to the total. 
The contribution from Type II 
supernovae is straightforwards ad it is incorporated in the Q-formalism.
The stellar ejecta are from \citet{Marigoetal1996,Marigoetal1998},
\citet{Portinari1998}, and \citet{Portinarietal1998} to whom we refer
for all details. The stellar lifetimes $\tau_M$ are from the tabulations by \citet{Bertellietal1994}
and take   the dependence of $\tau_M$ on the initial chemical
composition into account. Finally, for the purposes of this study we follow in detail only the 
total metallicity (the sum of the  abundances by mass of all elements heavier
than $\rm ^{4}He$), shortly  indicated by  $Z = \sum_{j > He} X_j$.

Last we write the equation for the current mass of a galaxy in form of stars $M_{s}(t)$: at any age $t$  
this is given by

\begin{equation}
  M_{s}(t) = M_{BM}(t) - M_g(t) 
\label{mass_stars}
\end{equation}
with obvious meaning of the symbols.

\subsection{The stellar initial mass function}
For this we choose the \citet{Salpeter1955}  law by number

\begin{equation}
 \phi(M)=  M^{-x}
\label{imf}
\end{equation}
\noindent
where $ x=2.35$. The IMF is normalized by choosing the fraction   $\zeta$ of stars more massive than
$M_n$,
i.e. the mass interval most contributing to chemical enrichment over the whole Hubble time

\begin{equation}
{ \zeta =  \frac{\int_{M_{n}}^{M_u}\phi(M){\times}M{\times}dM}
{\int_{M_l}^{M_u}\phi(M){\times}M{\times}dM}   }
\label{zeta}
\end{equation}

\noindent
Where $M_u$ and $M_n$ fixed and equal to   
$M_u =100 M_{\odot}$ and $M_{n} \simeq 1 M_{\odot}$,  the lowest mass limit   $M_l$ is left free. 
Following  \citet{Bressanetal1994,Tantaloetal1996,Tantaloetal1998} good choices force for $\zeta$ are
from 0.3 to 0.5 (and the values for $M_l$ are consequently derived).

\subsection{ The star formation rate}

The rate of star formation is assumed to depend on the gas mass according to

\begin{equation}
{ \Psi(t)= \frac{dM_g}{dt} = \nu M_{g}(t)^{k} }
\label{sfr}
\end{equation}
\noindent
where $\nu$  and $k$ are adjustable parameters.

The SFR normalized to ${ M_{BM}(T_G)}$ becomes

\begin{equation}
{ \psi(t)= \nu M_{BM}(T_G)^{k-1} G(t)^{k}  }
\label{sfr1}
\end{equation}
Linear and quadratic dependencies of the SFR on the gas
content, $k=1$ and $k=2$ respectively, were first proposed by  \citet{Schmidt1959}
and have been adopted ever since because of their simplicity
\citep[see][for a classical review of the subject]{Larson1991}. We adopt here $k=1$.

With the law of star formation of equation (\ref{sfr}), the resulting time dependence
of ${\rm \psi(t)}$  is driven by the rate of mass accretion onto the system. In
the closed-box description, the SFR is maximum at the
beginning, and since then  it continuously decreases  until galactic winds
occur. In the infall model, owing to the competition between the rate of gas
infall and gas consumption by star formation, the rate of star formation starts
small, increases to a maximum and then declines.
The age at which the peak
occurs, shortly indicated by $T_P$, approximately corresponds to the infall timescale $\tau$.

Finally, $\nu$ is the efficiency parameter of the star forming process.  Its  physical meaning is better
understood by  casting the SFR in a slightly different fashion. One can
identify $dt$ with the timescale $\tau$ of the mass accretion rate and assume $k=1$

\begin{eqnarray}
\Psi(t) = \frac{dM_g}{dt} = \nu M_{g}(t)^{k} \quad  \Rightarrow \quad
\frac{\Delta M_g}{M_g} = \tau \, \nu \quad  &&  \nonumber \\
\Rightarrow \quad \simeq  \frac{\Delta M_s} {M_{g}} = \tau \, \nu &&
\label{sfr_nor}
\end{eqnarray}

\noindent
where the ratio $\Delta M_s/M_{g}$ is the mass of gas already converted to stars with respect to the
mass of the
left-over gas. Furthermore, if the accretion $\tau$ is identified with the infall timescale $t_{ff}$
of a galaxy we may get rough estimates of the specific star formation efficiency. The infall timescale of
a galaxy can be approximated to the collapse timescale of primordial perturbations which depends on the
redshift but is independent on the galaxy mass.  Rough  estimates of $\tau$ yield values ranging
from 0.05 to 0.1 Gyr 
when the galaxy formation redshift is in the interval 20 to 1. 
This in turn implies $\nu \simeq$ 20 to 1 to assemble a typical $10^{12}\, M_\odot$ galaxy.
This efficiency
is lowered by a factor of ten at least if the mass assembly is diluted over the Hubble time.

According to  \citet{Cassara_etal2016} the shape of SFR as a function of time can
be schematically grouped according to the value taken by the ratio of the two parameters $\tau$ and
$\nu$ \citep[see Fig.2 of][]{Cassara_etal2016}.
With the above laws of gas accretion and star formation, they are able to model two main types of
objects: (i) in bulge-like models, characterized by high values
of $\nu$ and low values of $\tau$ (ratios $\tau/\nu \leq 0.1$), the SFR 
increases to a peak  on a relatively short timescale (on
average 0.5 Gyr), and soon after  declines. These models reproduce
the chemical pattern in the gas of early-type galaxies  at both low
\citep{Piovanetal2006a,Piovanetal2006b,Piovanetal2006c,PipinoMatteucci2011} and high redshift
\citep[e.g.][]{MatteucciPipino2002, PipinoMatteucci2011};
(ii) in the  disk-like models, characterized by low values of $\nu$ together with high values
of $\tau$ (ratios $\tau/\nu \geq 1$), the SFR  shows a slow rise followed by a slow decline. 
These models could well mimic disk and to some extent also irregular galaxies in the local Universe
\citep{Piovanetal2006a,Piovanetal2006b,Piovanetal2006c,Pipinoetal2013}.

Finally, we like to mention that a functional form for the SFR that could mimic the above systematic
variation with galaxy type (mass) is the so-called delayed exponentially declining law

\begin{equation}
\Psi(t) \propto  \frac{t}{\tau} exp(-\frac{t}{\tau} )\, .
\label{psidelay}
\end{equation}

\noindent
In this framework, the Schmidt law is the link between the gas accretion by
infall and the gas consumption by star formation. By varying the parameters $\tau$ and $\nu$ we
may model different types of galaxies  \citep{Buzzoni2002}.

Basing on these considerations and  taking the results of NB-TSPH simulations by
\citet{ChiosiCarraro2002}, \citet{MerlinChiosi2006,MerlinChiosi2007}, and \citet{Merlinetal2012}
as  reference templates
for the SFH in galaxies of different mass and  morphological types, we calculate
chemical models for different combinations of $\tau$ and $\nu$ that are meant to cover the whole
Hubble sequence of galaxies.

\subsection{Onset of galactic winds: energy feedback and gas heating-cooling }

Long ago \citet{Larson1974} postulated that the present-day Color-Magnitude Relations
of elliptical galaxies 
\citep[see][and references]{Boweretal1992,Kodamaetal1999,Kodamaetal2001,Terlevichetal2001}
could be the result of galactic winds powered by supernova explosions thus initiating a long
series of chemo-spectro-photometric models of elliptical galaxies standing on
this idea \citep[][and references therein]{Saito1979a,Saito1979b,MatteucciTornambe1987,ArimotoYoshii1987,
AngelettiGiannone1990, MiharaTakahara1994, Matteucci1994,
Gibson1998, GibsonMatteucci1997}.
In brief, gas is let escape from the galaxy and
star formation is supposed to halt when the total thermal energy of the gas
equates its gravitational binding energy.
The same scheme is adopted here in the models that take galactic winds into account, i.e. the term 
$[{d G_{i}(t) \over dt }]_{win}$ in eqn. (\ref{degas_i}) is at work.

The thermal energy of the gas is mainly due to  three contributions, namely Type Ia and
II supernovae and stellar winds from massive stars:

\begin{equation}
E_{th}(t) = \sum_{J} E_{th}(t)_J 
\label{Eth_tot}
\end{equation}

\noindent
where $J\equiv$ SNI for type Ia supernovae, $J\equiv$ SNII for type II supernovae, and 
$J\equiv W$ for stellar winds; each term has a similar dependence

\begin{equation}
E_{th}(t)_{J} = \int_{0}^{t} \epsilon_{J}(t-t')
            R_{J}(t') M_{BM}(T_{G}) dt'
\label{Esni_w}
\end{equation}

\noindent
with obvious meaning of the symbols.  The quantities 
$\epsilon_{J}(t-t')$'s and $R_{J}(t)$'s 
are the energies emitted by a supernova and/or stellar wind event and   
the   corresponding production rates, respectively. 
As the production rates are functions of the dimensionless variables
$G_{i}(t)$, the normalization factor $M_{BM}$  is required to calculate the energy in physical units.
The production rates  can be easily derived from the  equations governing the chemical evolution. 
The emitted energies incorporate  the cooling laws of supernova remnants and stellar winds by radiative 
cooling processes according to expression used by \citet[][]{Tantaloetal1998}.

Finally,  star formation and chemical enrichment of the model galaxies are halted,
and the remaining gas  is supposed to be expelled 
(winds) when the condition

\begin{equation}
E_{th}(t) \geq \left| \Omega_{g}(t)\right|
\label{eth_omg}
\end{equation}

\noindent
is verified.

\subsection{Gravitational potential of DM and BM}
To calculate the gravitational energy of the gas we make use of the analytical 
dynamical models  of \citet{Bertinetal1992}
and \citet{Sagliaetal1992} and adapt them to our case. DM is supposed to be 
already in situ, whereas the BM is supposed to fall into the gravitational 
well of 
the former and soon to reach the equilibrium configuration so that at each instant 
the description of \citet{Bertinetal1992}
and \citet{Sagliaetal1992} can be applied. 
In this description of galactic structure, the mass and radius of the DM, $M_{DM}$ and $R_{DM}$
respectively, are related to those of the BM,  $M_{BM}$ and
$R_{BM}$, by the relation

\begin{equation}
{  {M_{BM}(t) \over M_{DM} } \geq {1\over 2\pi} ({R_{BM}(t)\over R_{DM}})
       [1 + 1.37 ( {R_{BM}(t) \over R_{DM} } )]   }
\label{dark}
\end{equation}
\noindent
and the binding gravitational energy of the gas is given by

\begin{equation}
{ \Omega_{g}(t)=-{\alpha}_{BM} G {M_{g}(t) M_{BM}(t)\over R_{BM}(t) } -
G {M_{g}(t) M_{DM} \over  R_{BM}(t) } \Omega'_{BDM}  }
\label{gas_pot}
\end{equation}
\noindent
 where $G$ is the gravitational constant, ${\rm M_g(t)}$  the current 
value of the gas
mass, $\alpha_{BM}$  a numerical factor $\simeq 0.5$, and

\begin{equation}
{  \Omega'_{BDM}= {1\over 2\pi} ({R_{BM}(t)\over R_{DM}}) [1 + 1.37
            ( {R_{BM}(t) \over R_{DM} } )] }
\label{dark_pot}
\end{equation}
\noindent
the contribution to the gravitational energy given by the presence of DM. 
According to  \citet{Bertinetal1992} and \citet{Sagliaetal1992}, in equilibrium conditions 
 ${M_{BM}/M_{DM}} \simeq  {R_{BM}/R_{DM}}$.  With the  current estimates of $M_{DM}$ and $M_{BM}$ of the 
 $\Lambda$-CDM cosmogony both ratios are  equal to 0.16. With these values, the  factor 
 $\Omega'_{BDM}=0.04$  so that   the  total correction to the
gravitational energy of the gas (eq. \ref{gas_pot}) does not
exceed 0.3  of the term for the luminous mass.

\subsection{General remarks on the galactic models}

\textsf{Mass homology}. It is worth noting that with the above formalism, in absence of galactic winds
all the models are \textit{homologous in mass}  in the sense that the same solution (current
fractional gas and star mass) applies to galaxies
of different mass provided  suitably rescaled to the total asymptotic baryonic mass,
i.e. the  total baryonic mass reached at $t=T_G$. The same technique can also be used in presence 
of galactic winds by suitably rescaling the asymptotic
mass to the real value i.e. subtracted by the amount of gas mass definitely lost by the system in form of
galactic winds.

\textsf{Specific star formation rate}. It is also worth noticing that with above assumptions
the SFR in use is the specific star formation rate (SFR per unit baryonic mass, SSFR) which depends on three parameters, i.e.
$\nu$, $\tau$ and   $T_P$. Since $\tau$ and $T_P$ are correlated,
 each galaxy model here is characterized here only by the parameters $\nu$ and $\tau$.

\textsf{Groups of galaxy models}. For the purposes of this study we have calculated
 three groups of models, labeled  A, B and C:

\textit{Group A}.  In the
models of group A, we assume that all galaxies begin their evolutionary history at redshift $z$=20
and suppose that the mass
accretion timescale $\tau$ corresponds to the free-fall timescale for the $\rho_{200}$ over-density of
the proto-galaxy with respect to the surrounding medium at this value of the redshift.
The free-fall timescale is given by

\begin{equation}
t_{ff}(z) =  \sqrt{ \frac{3\pi} {32 G \rho_{200} }  }
\label{free_fall}
\end{equation}

\noindent
for the homologous collapse of a homogeneous sphere of gas. This timescale is the same for all  galaxies
independently of their mass. The free-fall timescale $t_{ff}(z)$ goes from about $3.5 \times 10^7$ yrs 
at z=20 to about $1.0\times 10^9 $ yrs at $z=1$.

For the star formation efficiency parameter $\nu$ we adopt the
constant value $\nu=10$.  Owing to the very short mass accretion timescale and high value of $\nu$
these models are very similar to the ideal situation of the closed-box approximation.
The baryonic mass of the models spans the range $10^7$ to $10^{12}\, M_\odot$.

\textit{Group B}. At the light of the considerations on the type of SFH in real
galaxies, in models of
group B we adopt values for the accretion timescale $\tau$ ranging
from 6 Gyr to  2 Gyr as the baryonic mass increases from $10^7\, M_\odot$ to
$10^{12}\, M_\odot$, but keep unchanged the value for the parameter $\nu$, i.e. $\nu=10$ for all the models.
Since the ratio  $\tau/\nu$ of the models of groups A and B is always lower than one, they are
best suited to represent early type objects of any mass going from dwarfs to bulges and massive
ellipticals \citep{Cassara_etal2016}.

\textit{Group C}. Finally, the models of Group C more closely follow the classification
by \citet{Cassara_etal2016} who
rank the galaxy SFR by means of the ratio $\tau/ \nu$. First of all,  we stretch the interval of
the accretion
timescale $\tau$ assigned to each BM mass. It goes now from 2 Gyr for the
$M_{BM}=10^{12}\, M_\odot$
galaxy to 10 Gyr for the $M_{BM}=10^{7}\, M_\odot$. Second, for each value of $\tau$ we
explore three values of
$\nu$, namely $\nu=0.1$, 1, and 10. In this way we can model galaxies  along the whole Hubble sequence
by varying the ratio $\tau/\nu$ depending on the galaxy  mass: low values for the most massive ones
corresponding to
the massive early types,  intermediate values for the less massive ones (going from
intermediate early type  to massive
spirals), and high values for the less massive galaxies like low spirals and irregulars.
Group C partially overlaps group B.

\begin{table*}
\begin{center}
\caption{Parameters adopted in the chemical models. Masses are in $M_\odot$, the
timescale $\tau$  is in Gyr, and the radii $R_{BM}$ and $R_{DM}$ are in kpc. }
\label{parchem}
\begin{tabular}{llrrrrrrrrrrrr}
\hline
\multicolumn{1}{l}{$M_{BM}$}   &
\multicolumn{1}{l}{$M_{DM}$}    &
\multicolumn{1}{c}{$R_{BM}$}    &
\multicolumn{1}{c}{$R_{DM}$}    &
\multicolumn{1}{c}{$\zeta$}     &
\multicolumn{1}{c}{$\tau$}     &
\multicolumn{1}{c}{$\nu$}       &
\multicolumn{1}{c}{$\tau/\nu$}  &
\multicolumn{1}{c}{$\tau$}     &
\multicolumn{1}{c}{$\nu$}       &
\multicolumn{1}{c}{$\tau/\nu$}  &
\multicolumn{1}{c}{$\tau$}     &
\multicolumn{1}{c}{$\nu$}       &
 \multicolumn{1}{c}{$\tau/\nu$} \\
 \hline
\multicolumn{5}{c}{    }        &
\multicolumn{3}{c}{Group A}    &
\multicolumn{3}{c}{Group B}    &
\multicolumn{3}{c}{Group C}    \\
\hline
          &                   &      &       &     &      &      &        &   &      &     & 10.0 & 0.1 & 100   \\
$10^{7} $ & $6\times 10^{7}$  & 0.13 &  1.35 & 0.3 & 0.01 & 10.0 & 0.003  & 6.0 & 10.0 & 0.6 & 10.0 & 1.0 & 10 \\
          &                   &      &       &     &      &      &        &   &      &     & 10.0 &10.0 &  1  \\
\hline
          &                   &      &       &     &      &      &        &   &      &     &  8.0 & 0.1 & 80 \\
$10^{8} $ & $6\times 10^{8}$  & 0.28 &  2.90 & 0.3 & 0.01 & 10.0 & 0.003  & 5.0 & 10.0 & 0.5 &  8.0 & 1.0 & 8 \\
          &                   &      &       &     &      &      &        &   &      &     &  8.0 &10.0 &0.8 \\
\hline
          &                   &      &       &     &      &      &        &   &      &     &  6.0 & 0.1 & 60 \\
$10^{9} $ & $6\times 10^{9}$  & 0.61 &  6.26 & 0.3 & 0.01 & 10.0 & 0.003  & 4.0 & 10.0 & 0.4 &  6.0 & 1.0 & 6 \\
          &                   &      &       &     &      &      &        &   &      &     &  6.0 &10.0 & 0.6\\
\hline
          &                   &      &       &     &      &      &        &   &      &     & 4.0 & 0.1 & 40   \\
$10^{10}$ & $6\times 10^{10}$ & 1.32 & 13.48 & 0.3 & 0.01 & 10.0 & 0.003  & 3.0 & 10.0 & 0.3 & 4.0 & 1.0 & 4 \\
          &                   &      &       &     &      &      &        &   &      &     & 4.0 &10.0 & 0.4   \\
\hline
          &                   &      &       &     &      &      &        &   &      &     & 3.0 & 0.1 & 30   \\
$10^{11}$ & $6\times 10^{11}$ & 2.85 & 29.04 & 0.3 & 0.01 & 10.0 & 0.003  & 2.5 & 10.0 & 0.2 & 3.0 & 1.0 & 3 \\
          &                   &      &       &    &      &      &        &   &      &     & 3.0 &10.0 & 0.3   \\
\hline
          &                   &      &       &     &      &      &        &   &      &     & 2.0 & 0.1 & 20   \\
$10^{12}$ & $6\times 10^{12}$ & 6.13 & 62.53 & 0.3 & 0.01 & 10.0 & 0.003  & 2.0 & 10.0 & 0.2 & 2.0 & 1.0 & 2 \\
          &                   &      &       &     &      &      &        &   &      &     & 2.0 &10.0 & 0.2   \\
\hline
\end{tabular}
\end{center}
\end{table*}

It goes without saying that other combinations of two out of the
three parameters ($\tau$, $\nu$, and hence $\tau/\nu$) are possible. Since the aim here is to calculate
galaxy models whose SFR
and SFH  closely resemble the ones observed in real galaxies, the choice we have
made is fully adequate to our purposes.

Finally, we like to point out that the models of
Group A are meant to represent a sort of reference sample corresponding to the ideal situation
of the closed-box  approximation. They will be used only to evaluate the effects on the SFRD(z) of an
exponentially decreasing rate of star formation.

Table \ref{parchem} lists all the parameters adopted for the
chemical models in usage.

\subsection{Results for the chemical models}

The model galaxies are calculated from $z_{f}=20$ to $z=0$ i.e. from the rest-frame
age $t=0$ Gyr to the maximum age of $T_{G}=13.75$ Gyr where  $T_G=t_u(z=0) - t_u(z=z_f)$
with $t_u(z)$ being  the age of the Universe for the
adopted cosmological model.

If for any reason,
we need to change the redshift of galaxy formation from $z_f$ to $z_{f}^{*} \le z_f$
(keeping unchanged all other input parameters) 
the same models can be used  provided their rest-frame age is simply
limited to the interval from $t=0$ at
$z_{f}^{*}$ to
$T_{G}^{*}= T_{G,(z_{f}=20)} - t_u(z_{f}^{*})$, where $t_u(z_{f}^{*})$ is the age of the universe
at $z=z_{f}^{*}$.
In other words,  at $z=0$
the new galaxy is younger than the previous one.

On purpose, in this first step of the analysis,  the occurrence of galactic winds is not considered.
This means that the
energy input from Type II
and Type Ia supernovae and galactic winds are turned off so that star formation can occur
 all over the  life of galaxies.

Finally, the discussion below is limited to the models of case B. Those of cases A and C have similar 
trends and behaviour.

\begin{figure*}
\centering{
{\includegraphics[width=8.0cm,height=8.0cm]{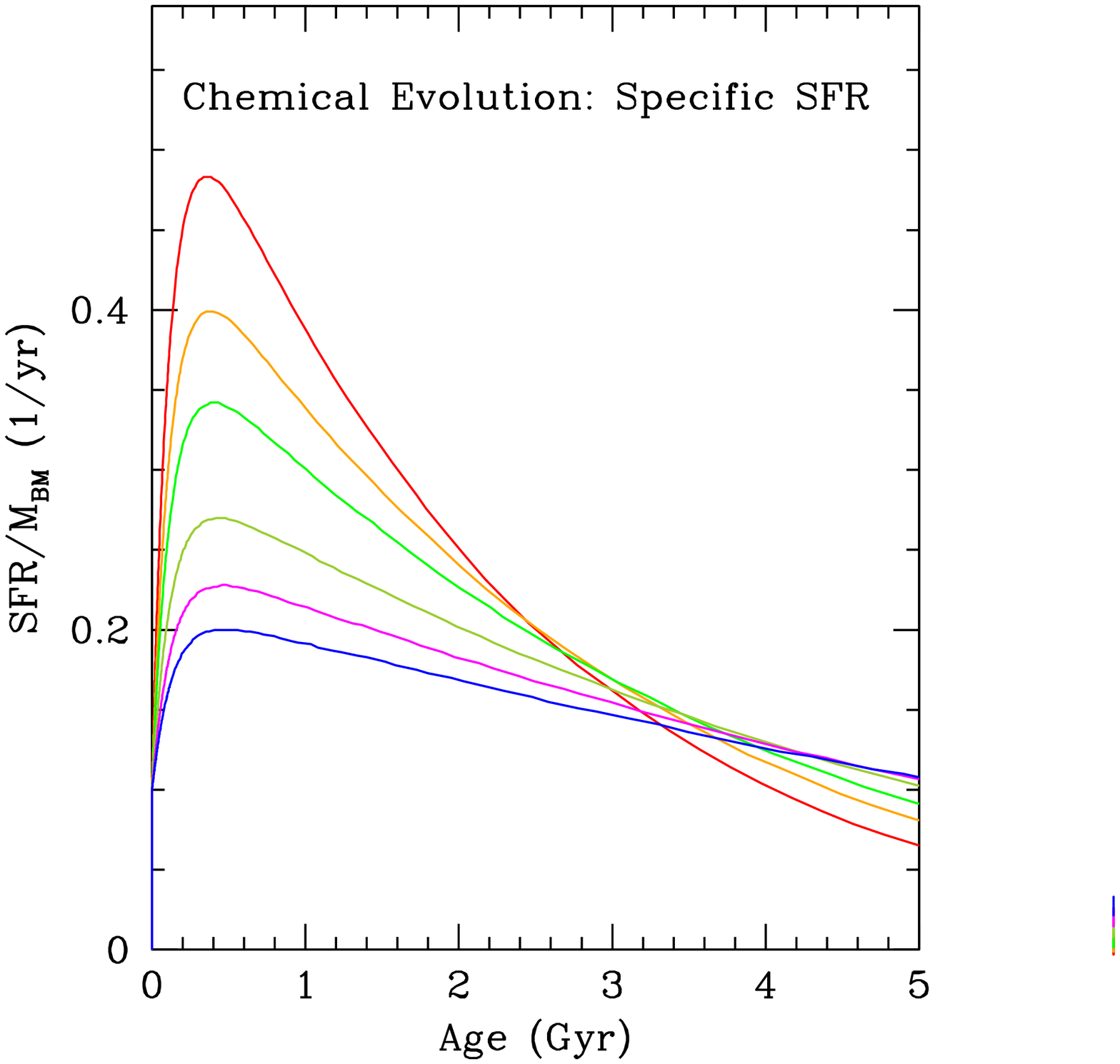}    }
{\includegraphics[width=8.0cm,height=8.0cm]{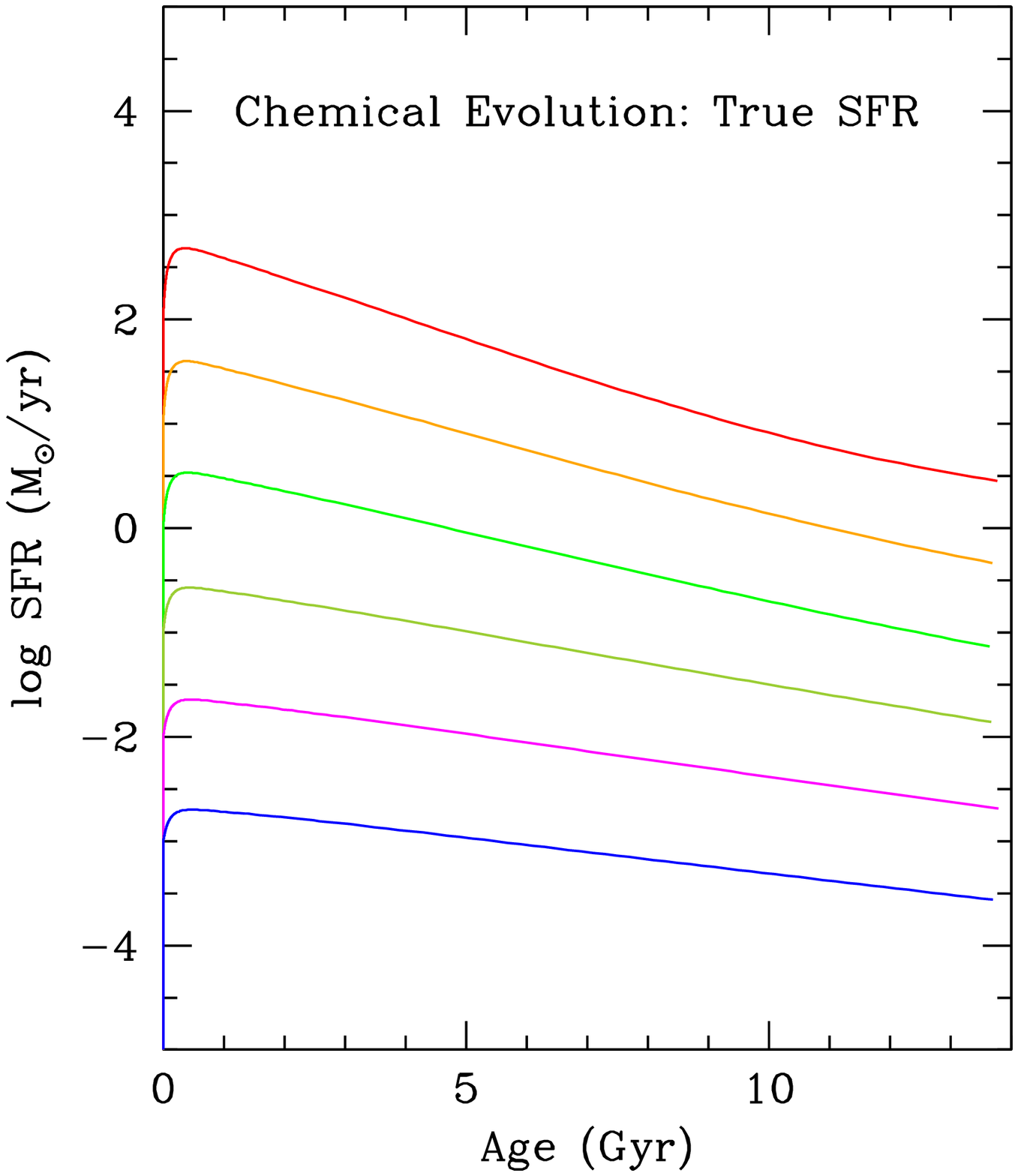}         }   }
\caption{\textsf{Left  Panel}: The SSFR of models B in $ yr^{-1}$
for the galaxies of different $M_{BM}$, different accretion (collapse) timescale $\tau$ and
efficiency $\nu = 10$.  The mass $M_{BM}$ increases from $10^7$ to $10^{12}\, M_\odot$
from the bottom to the top.
No galactic winds are supposed to occur. The time is the age of the
galaxy in the rest-frame.
\textsf{Right Panel}:
The same as in the left  panel but for the true SFR in units of $M_\odot \, yr^{-1}$.  }
 \label{sfr_models_B}
\end{figure*}

\textsf{Star formation}. The specific (in units of $yr^{-1}$) and true star
formation (in $M_\odot\, yr^{-1}$) of
the model galaxies are shown in the left and right panels  of
Fig. \ref{sfr_models_B}. 
As expected, the SSFRs look very similar to each other, whereas the
true rates may significantly
change with the galaxy mass. From now on, different values of $M_G$ are identified in all figures with the
following colors: blue ($10^{7} M_{\odot}$),
magenta ($10^{8} M_{\odot}$),
olive green ($10^{9} M_{\odot}$),
green ($10^{10} M_{\odot}$),
orange ($10^{11} M_{\odot}$) and
red ($10^{12} M_{\odot}$).

\begin{figure}
\centering{
{\includegraphics[width=7.5cm]{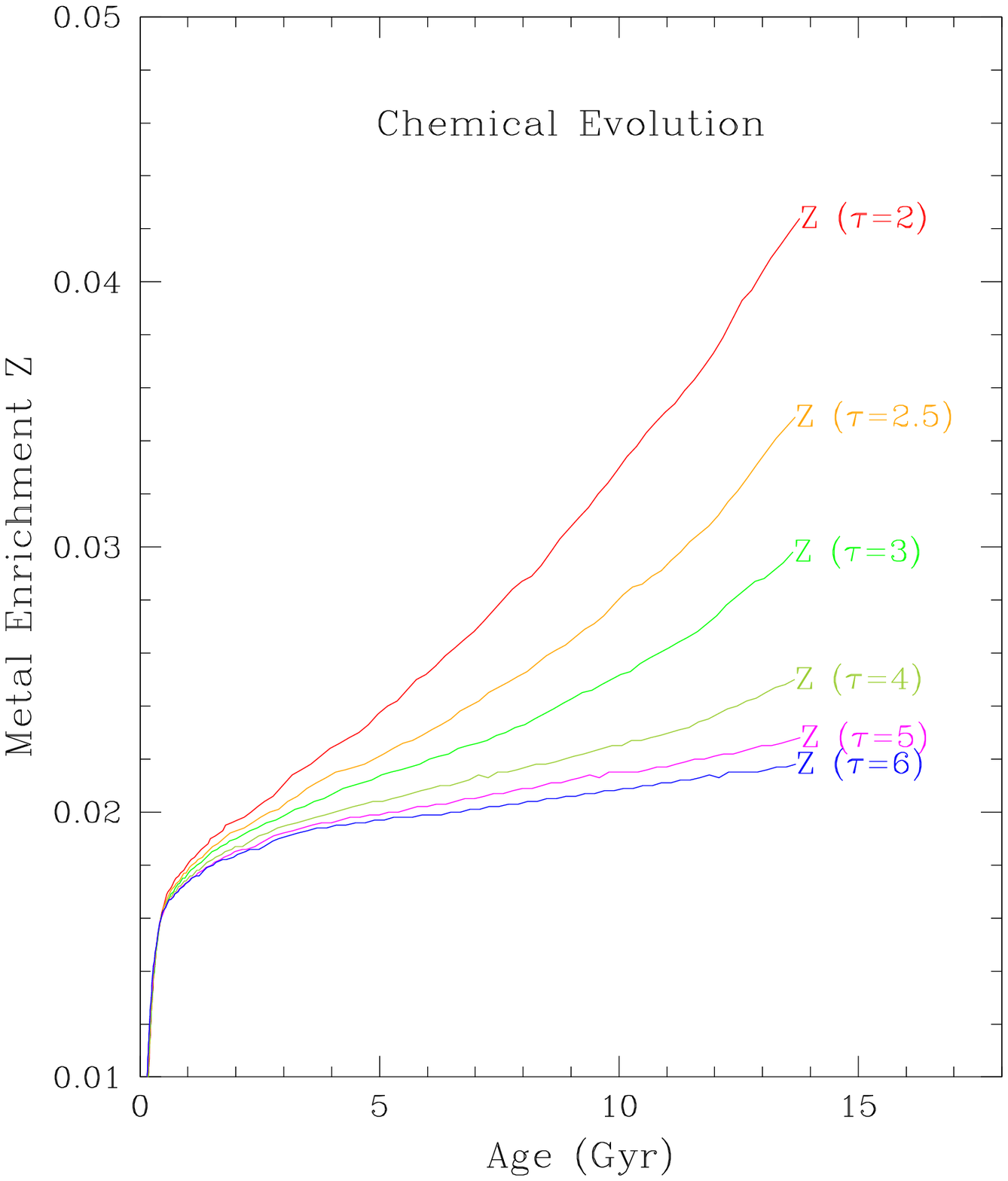}  } 
}
\caption{
The metallicity vs time relation
for the galaxies of group B with different $M_{BM}$, different accretion  timescale $\tau$ and
efficiency $\nu = 10$. No galactic winds are supposed to occur. The mass $M_{BM}$ incresases from $10^7$ to $10^{12}\, M_\odot$
from the bottom to the top.
}
 \label{met_models}
\end{figure}

\textsf{Metallicity}.
The temporal variation of the metallicity $Z$ for the model galaxies is shown in
 Fig. \ref{met_models}.
Owing to the rather high value of $\nu$ and parameter $\zeta$ of the IMF normalization,
high metallicites are built up
in the galaxies. This is less of a problem because by lowering $\nu$ and/or $\zeta$
one would obtain similar
results but lower
values of the  metallicity at the present time without changing the overall behavior of the solution.

\begin{figure}
\centering{
{\includegraphics[width=7.50cm]{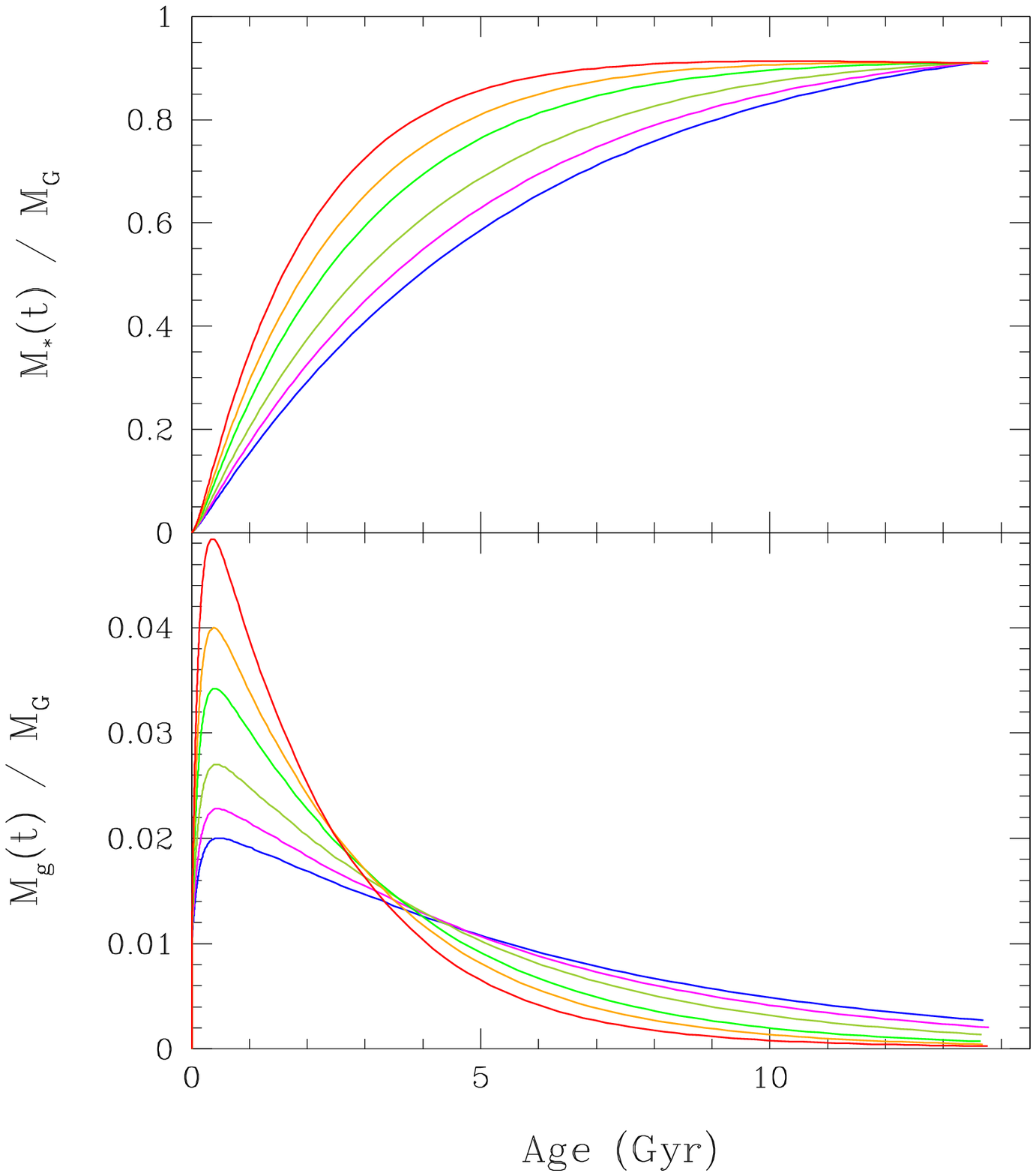}  }
}
\caption{
The gas and star content  vs time relationships
for the galaxies belonging to group B withdifferent $M_{BM}$, different accretion
timescale $\tau$ and efficiency $\nu = 10$.
The mass $M_{BM}$ increases from $10^7$ to $10^{12}\, M_\odot$
from the bottom to the top.
}
 \label{gas_stars}
\end{figure}

\textsf{Gas and Star contents}.
Finally, in Fig.\ref{gas_stars} we show the temporal variation of the fractional masses of gas
(bottom panel) and stars (top panel) for the models of group B. The timescale of mass
accretion varies with the galaxy mass as reported in Table \ref{parchem}. The intrinsic efficiency
of star formation  is the same for the models on display (i.e. $\nu$=10).   
Because of the high efficiency of star formation,  all the models
have the peak of activity within the first Gyr of their lifetime.

\textsf{The role of $\nu$}. Concluding this section, it is worth commenting on the role of the intrinsic
efficiency $\nu$ in shaping the final time-dependence of the SFR in infall models
of galaxy formation and evolution. So far the discussion of the results for groups B and C has been limited
to models with efficient SFR represented here by all the cases with $\nu=10$.
The peak of activity is always confined within the first Gyr. 
Clearly for $\nu=10$ case C does not differ too much from case B.
We give the preference to this particular choice for the
parameter $\nu$ in view of
the discussion below concerning the SFRD(z).

\begin{figure}
\centering{
{\includegraphics[width=7.5cm]{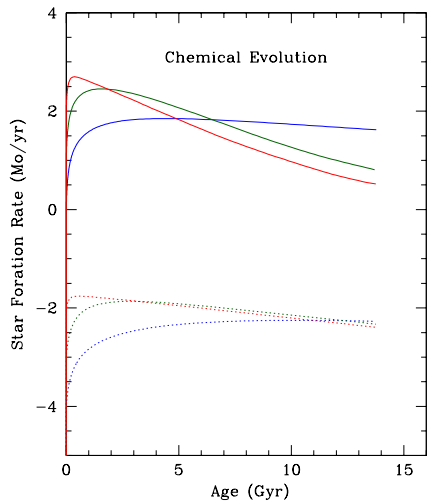}  }     }
\caption{ The SFR versus time relationship
for the galaxies of different $M_{BM}$ belonging to group C at varying $\tau$ and $\nu$. Two values of
the galaxy
mass (baryonic component) are considered, namely $10^{8} M_\odot$ and $10^{12} M_\odot$. The
values of $\tau$ and $\nu$ are listed in Table \ref{parchem}.
}
 \label{nu_tau_C}
\end{figure}

In any case it is worth emphasizing that the role of $\nu$ is of paramount importance in shaping
the overall time-dependence of the SFR. The situation is best illustrated in
Fig. \ref{nu_tau_C} that displays the SFR vs time of the $1\times10^{12} M_\odot$ and $1\times10^8 M_\odot$
galaxies of Group C (three values of $\nu$ for each case). The models gradually change their SFR from early
peaked to ever-continuing according to the value of the ratio $\tau/\nu$, in other words along the Hubble
sequence of galaxies passing from early types (low ratios $\tau/\nu$) to disk-like objects
(intermediate ratios $\tau/\nu$), and finally to irregulars (high ratios $\tau/\nu$). This trend of
the star formation was suggested by  \citet{Sandage1986}, examining the SFR in galaxies of different types,
and  more recently confirmed by studies of SFHs based on absorption line indices by
\citet{Thomasetal2005}, and NB-TSPH numerical models of galaxy formation and evolution
\citep{ChiosiCarraro2002,Merlinetal2012}.

\subsection{Remarks on the star formation rate}

As already mentioned, the time dependence of our SFR is the delayed exponential law,
see eq. (\ref{psidelay}), which is implicit to the galactic chemical models with gas accretion.
The reasons why the simple exponential law,
\begin{equation}
\Psi(t) \propto \frac{1}{\tau} exp(- \frac{t}{\tau}) ,
\label{sfr_delay}
\end{equation}

\noindent
adopted long ago by  \citet{Tinsley1972} and in usage even today, must be abandoned
have been discussed many times (see the classical studies by \citet{LyndenBell1975},
\citet{Chiosi1980}
and the recent review by
\citet{Matteucci2016}  so that they are not repeated here).

In favor of the time delayed exponential law were the original models by \citet{Chiosi1980} and the long
list of studies dedicated to the evolution of chemical elements in galaxies of different morphological
type, going from bulges and early-type objects to disk  and even irregular galaxies
\citep[see][for exhaustive reviews and referencing of this issue]{Matteucci2012,Matteucci2016}.
In support to the delayed exponential law is also the study of \citet{Gavazzietal2002} with the
spectro-photometric data of galaxies in the Virgo cluster.

These classical analytical representations of the SFR have been
recently questioned by \citet{Oemleretal2013}  basing their  analyses on the SSFR in the redshift
interval $z\leq 1$.  They concluded
that the standard laws cannot explain both the tail of high specific SFR at z=1 and the low value 
we see today at
$z=0$. They also argued that the starbust hypothesis cannot solve the problem. \citet{Gladdersetal2013}
argue that a log-normal SFH  of galaxies successfully describes both the SFRD over cosmic times
and the present-day distribution of the SSFR of galaxies
 and the evolution of this quantity up to $z\simeq 1$. The log-normal SFR law they assume is

\begin{equation}
SFRD(t) \propto  \frac{1}{t \tau} exp[ - \frac {(ln(t) - T_0)^2} {2\tau} ]
\label{sfr_logno}
\end{equation}

\noindent
where $T_0$ and $\tau$ (not to be mistaken with the timescale of gas accretion in galaxies)
are the cosmic SFRD's half mass time and width [in units of ln(time)]. 
Basing on  the notion that log-normal
laws seem to be  ubiquitous in Nature \citep{Limpert2001},  they take the
SDSS sample of local galaxies (2094 objects),  assign them  a log-normal SFR, and derive
for each object the SFR (i.e. the parameters $T_0$ and $\tau$), while ensuring that the ensemble
of SFRs summed to the SFRD.
\citet{Abramsonetal2016} go ahead along this line of reasoning. Adding and folding together a
large number of log-normal SFHs parameterized by $T_0$ and $\tau$, they argue that this
simple minded model
reproduces (i) the stellar mass functions at $z\leq 8$; (ii) the slope of the SFR vs stellar mass
relation (the Galaxy Main Sequence) at $z \leq 6$; (iii) galaxy downsizing; (iv) and a correlation
between the formation timescale and the SSFR(M$_s$,t).

In our view, the straight inference of a log-normal SFR in single galaxies contributing to the total
cosmic SFRD  is somewhat arbitrary and misleading. 
The cosmic SFRD is not the simple summation
of that of many galaxy SFRs because each galaxy may differ from the others  nor all types of
galaxy occur in equal number, but it results from the number weighted summation of many objects of
different type and SFHs (see below). The SFR of a galaxy might not be lognormal and yet
the cumulative
effect of many of them may turn out to look as a log-normal distribution.
 For these reasons we prefer to describe the SFR
of galaxies as independent entities with the time-delayed law.

\section{ Third building block: photometry  }\label{third_block}

The integrated monochromatic flux generated by the  stellar content of a galaxy
of age $T$ is defined as

\begin{equation}
F_{\lambda}(T) = \int_0^T \Psi[t,Z(t)]\ {\it sp}_{\lambda}[\tau',Z(\tau')]\ dt
\label{popsynt1}
\end{equation}

\noindent
where $\Psi[t,Z(t)]$ is the SFR at the current age $t$ and metal content $Z$ (chemical 
composition in general), $ {\it sp}_{\lambda}[\tau',Z(\tau')]$ the integrated monochromatic  flux of single 
stellar population (i.e. of a coeval, 
chemically homogeneous assembly of stars, shortly named SSP)  with age $\tau'$ and metallicity $Z(\tau')$, and finally 
$\tau'=T-t$. The flux of a SSP is in turn given

\begin{equation}
{\it sp}_{\lambda}[\tau',Z(\tau')] =
\int_{M_l}^{M_u} \phi(M)\ f_{\lambda}[M,\tau',Z(\tau')]\ dM
\label{popsynt2}
\end{equation}
\noindent
where $\phi(M)$ is the stellar IMF and  
$f_{\lambda}(M,\tau',Z)$ the monochromatic flux of a star of mass $M$,
metallicity $Z$, and age $\tau'=T-t$. M$_l$ and M$_u$ respectively,
define the mass range within which stars are generated by each event of
star formation.
\noindent
The metallicity dependence 
of the rate of star formation $\Psi(t,Z)$ is customarily neglected and equally for the time and 
metallicity 
dependencies of the IMF.

The flux of a SSP, ${\it sp}_{\lambda}(\tau',Z)$, is calculated by
integrating equation (\ref{popsynt2}) along an isochrone of age $\tau'$ populated  by ``virtual stars'' 
with luminosty $L$, effective temperature $T_{eff}$, mass $M$,  age $\tau'$, and composition $Z$.
For any star along an isochrone, the relations
connecting luminosity, effective temperature, and age
are derived from the library of stellar models, while the
 flux $f_{\lambda}(M,\tau',Z)$ emitted by such a star is obtained
 from the library of stellar spectra. Sources of stellar tracks, isochrones, spectra, and  SSPs
in different photometric systems are from
\citet{Bertellietal2008,Bertellietal2009}.

\textsf{ Cosmological evolution of magnitudes and colours}.
In the course of this study, we need the magnitudes and colors of the galaxies not only in the 
rest-frame but also as function of the redshift.  Following \citet{GuiderdoniRoccaVolmerange1987},
the apparent magnitude of a galaxy at redshift $z$ in a pass-band $\Delta\lambda$    is

\begin{equation}
 m(z)=(m-M)_{bol}(z)+K(z)+E(z)+M(0,t_0)
\end{equation}

\noindent
where $K(z)$ and  $E(z)$ are the cosmological and evolutionary corrections

\begin{equation}
 K(z)= M(z,t_0) - M(0,t_0)  
\end{equation}

\begin{equation}
 E(z)= M(z,t_z) - M(z,t_0)
\end{equation}

\noindent
and where $M(0,t_0)$ is the absolute magnitude in the pass-band
$\Delta\lambda$
derived from the rest-frame spectrum of the galaxy at the current time,
$M(z,t_0)$  is the absolute magnitude in the pass-band
$\Delta\lambda$ derived from the spectrum of the galaxy at the current time
but redshifted at $z$, and
$M(z,t_z)$ is the absolute magnitude in the pass-band
$\Delta\lambda$ obtained from the spectrum of the galaxy at the time
t$_z$ and redshifted at $z$.

\begin{figure}
\centering{
{\includegraphics[width=7.5cm,height=8.0cm]{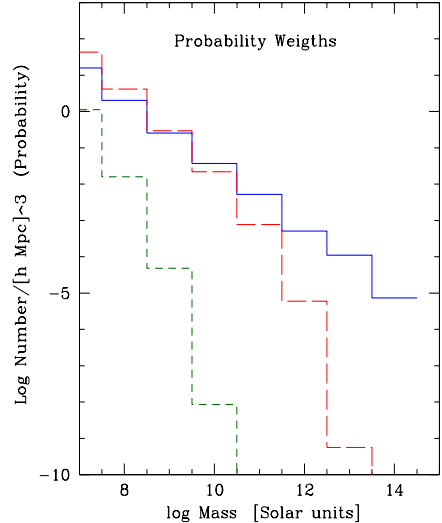}  }  }
\caption{ The expected number of DM halos
as a function of the halo mass $M_{DM}$ at
three selected values of the redshift, namely z=20 (short dashed line in dark green), z=6
(long dashed line
in red), and z=0 (solid blue line).     }
\label{lukic_mass_histo}
\end{figure}

\section{The Cosmic Star Formation Rate from theory }\label{SFRD_theory}

It is worth emphasizing from the very beginning that in the course of the analysis
and companion discussion
we will use two mass scales: (i)  The scale of the halo masses, i.e. we will refer to galaxies by their
halo mass which is roughly coincident with the total mass ($M_{G} \equiv M_{DM} + M_{BM}
\simeq M_{DM}$) to determine the number of halos per unit volume as a function of the halo mass and
redshift. (ii) The scale of the baryonic component hosted by a halo, made of gas and stars.
This scale will be used to read from the sample of chemical models their  SFR  and photometric
properties (magnitudes and colors) as a function of the
BM mass, age, redshift, etc. The relationship between the two mass scales is given in the first two
columns of Table  \ref{parchem}.

Second, we will use the grids of models of the group B, choosing the one appropriate to the mass halo,
according to
the mass scale  $M_{BM} \simeq M_{DM}/6$.  We have already described these models in the previous section.
However, we recall here that  for each model we know
both the SFR and the SSFR,
the abundance of  metals $Z(t)$ (for the present aims the
total metallicity
is fully adequate), the mass in gas $M_g(t)$ and the mass in stars $M_s(t)$,
the integrated magnitudes in the pass-bands $M_{\Delta\lambda}$ of the
Johnson-Cousins and/or
HST-WFPC photometric systems, the cosmological evolution of these magnitudes, i.e.
the $K_{\Delta\lambda}$ and
$E_{\Delta\lambda}$ corrections as a function of the redshift.

It is worth recalling here that these models not only fit the main average properties of the galaxies
in the local and distant Universe, see for instance \citet{Bressanetal1994}, \citet{Tantaloetal1998},
and \citet{Tantaloetal2010}, but also their SFHs agree  with the
results from NB-TSPH simulations of galaxy
formation in cosmological context and according to the so-called early hierarchical
scheme \citep{ChiosiCarraro2002, MerlinChiosi2006,
MerlinChiosi2007, Merlinetal2012}.
Therefore, the simple infall models presented here
can be safely used to study the mean properties of galaxies in the context
of the early hierarchical view of galaxy formation and evolution.

\subsection{Distribution of halos in number and mass}
We start the analysis by looking at  the mass distribution
of the DM halos at each value of the redshift. This is simply derived as the number
of DM halos within a small interval $\Delta z$ centered at  few selected values of the
redshift ($\Delta z = 0.02$). The results are listed in Table
\ref{count_lukic} and are plotted in Fig.\ref{lukic_mass_histo}.
The visual inspection of Fig. \ref{fig_lukic} yields a qualitative estimate
of the maximum value of the mass distribution at each redshift, which means
that DM halos with mass in excess of this value have such a low probability
of occurrence that they can be neglected   for any practical purpose.
The histograms of Fig.\ref{lukic_mass_histo} shows
the comoving number density of halos as a function of the halo mass for three selected values
of the redshift, namely $z$=20 (short dashed line), $z$=6 (long dashed line) and $z$=0 (solid line).
The mass distribution for all other values of redshift can be easily derived from the entries of Table
\ref{count_lukic}.
It is worth calling the attention on the steeper decrease of the number of halos at increasing halo mass
and increasing redshift. While for $z$=0  each  step of the histogram   roughly decreases
by the same amount, this is not the case for $z$=6 and higher in which the steps decrease more and more
 at increasing halo mass. This behavior of the number frequency distribution will have
far reaching consequences.

\begin{figure}
\centering
\includegraphics[width=7.5cm]{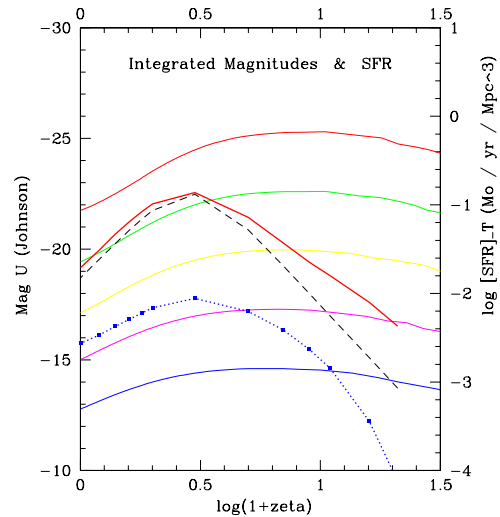}
\caption{ Three groups of data are displayed: (i) the integrated absolute U magnitude of the
model galaxies  as a function of the redshift (on logarithmic scale); each galaxy is indicated
by a solid line with different color code according to the mass of the BM component. The galaxies 
on display
have BM masses of $10^8$ to $10^{12} M_\odot$ from the bottom to the top of the Figure 
(blue, magenta, yellow, green, red).
(ii) The total magnitude in the U-pass-bands of
all the galaxies present in the ideal sample of  $1\, Mpc^3$ volume according to the
\citet{Lukicetal2007} statistics (the blue dotted line with filled circles).
The y-axis for the magnitudes is at left hand side of the panel.
(iii) The SFRD(z)
for the same sample of galaxies (the red solid line). Finally the analytical fit of the
\citet{MadauDickinson2014} SFRD
(the black dashed line).
The y-axis for the SFRD is at right hand side of the figure.}
\label{Umag_gal}
\end{figure}

\subsection{The reference case for the SFRD(z)}

In Fig. \ref{Umag_gal}  we present three groups of data:

(i) The integrated absolute U magnitude of the
model galaxies  as a function of  redshift (on logarithmic scale); each galaxy is indicated
with a different color code. As expected the absolute magnitude  first decreases
(the luminosity increases) and then increases (the luminosity decreases) as the redshift decreases toward
zero. The peak in the luminosity occurs when the rate of star formation is maximum.

(ii) The total luminosity and total magnitude in the U-pass-bands of
all the galaxies present in the ideal sample contained in  $1\, Mpc^3$ according to the
\citet{Lukicetal2007} statistics. The total U flux is  given by

\begin{equation}
[F_{\Delta\lambda}(z_j)]_T  = \sum_i N_{i}(M_{DM}, z_j) F_{{\Delta\lambda}, i} {\Delta z_{j}}
\end{equation}
where $[F_{{\Delta\lambda}, i}]_T$ is the flux in the chosen pass-band of the generic
galaxy  $i$ of mass $M_{G,i} = M_{BM,i} + M_{DM,i}$, and $\Delta z_j$ is the generic redshift 
range centered
on $z_j$ and defined by  $0.5\times[z_{j} - z_{j-1}]$ and $0.5\times[z_{j+1} - z_{j}]$.
The number of galaxies $N_{i}(M_{DM,i}, z_{j})$ at the
generic redshift $z_j$ is calculated using the mass scale of the DM halos, whereas the photometric
properties are obtained using the mass scale of BM and more precisely the mass in stars
existing in the galaxy at the time $t$ or redshift $z$. The indices $i$ and $j$ runs over the 
whole grids of
 masses and redshifts under considerations. 
The total U magnitude is shown by the dotted blue line in Fig. \ref{Umag_gal}. The magnitude
 scale along the y-axis is on the left hand side of the figure.

(iii) Similar procedure is applied to derive the total true SFR for the galaxies in
the same ideal sample contained in  a volume of $1 \, \rm Mpc^3$.

\begin{equation}
[SFRD(z_j)]_T  = \sum_i N_{i}(M_{DM}, z_{j} ) SFR_{i}(z_{j}) {\Delta z_{j}}
\label{sfr_N}
\end{equation}
where the indices $i$ and $j$ run over the whole mass and redshifts grids as before.
The SFRD(z)  is the red solid curve. The
scale for the   SFR is along the y-axis at the right hand side of the
figure. The SFRD(z) has the same trend of the total U-magnitude, thus confirming that
the UV light is a good tracer of star formation.

\begin{figure}
\centering{
{\includegraphics[width=7.5cm]{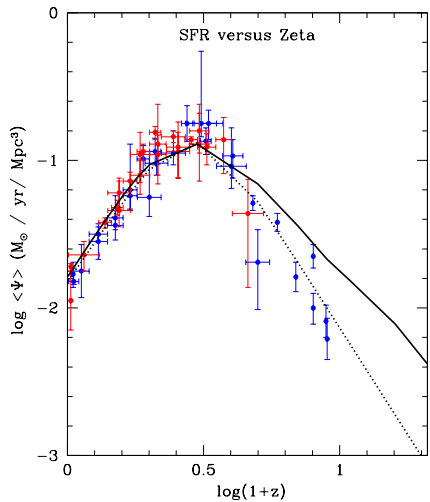}} }
\caption{The theoretical SFRD(z) predicted from galaxy  models of group B (solid black line) compared with
the observational data (blue and red filled circles with error bars) and the analytical
fit of   \citet{MadauDickinson2014} (dotted line).}
\label{sfr_madau}
\end{figure}

The results are shown in Fig.\ref{sfr_madau} and compared with the data and the empirical best-fit
relation of \citet{MadauDickinson2014} given by eqn. (\ref{eq_madau_dickinson}) (the dotted black line).

Theory and data nearly agree
in the location of the peak (redshift $z \simeq 2$) and at
the low redshift side  (descending branch), whereas they may differ up to a factor of 
three beyond the peak towards the past.
The provisional conclusion we could derive at this stage is that
the theoretical  SFRD(z)    and the data of \citet{MadauDickinson2014}
 fairly agree each other,
thus indicating that our simple model of the cosmic SFR
well reproduces the  observational data.

It is worth emphasizing here that for each bin of redshift, the SFRD(z) is obtained by
summing up the contribution from galaxies of different mass, and in particular different history
and stage of star formation. For instance
in the redshift interval $1 < z < 3$  we may have both galaxies with increasing  SFR and galaxies with
descending SFR. The change in the slope of the SFRD(z) at $z\simeq 2$ implies a change
in the slope of the mean SFR in the galaxy population. At $z > 2$ galaxies with
increasing SFR dominate, the opposite at  $z < 2$, they balance each other at $z \simeq 2$.
The SFRD(z) does not tell the behavior of individual galaxies, but only the current
mean behavior of whole galaxy  population (see also the discussion in Section \ref{dissect} below).

\begin{figure}
\centering{
{\includegraphics[width=7.5cm]{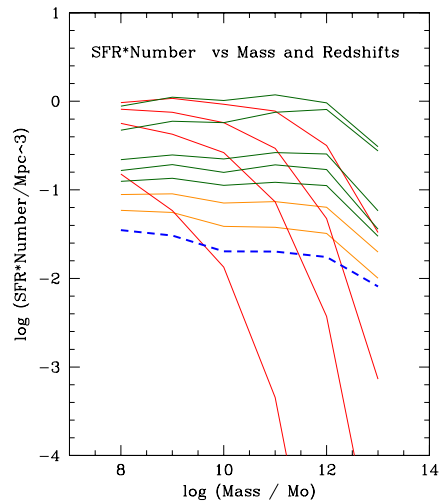} } }
\caption{Contribution to the  SFRD(z) from galaxies of different mass at varying
redshifts. The red
lines are for redshifts $z$=20, 15, 10, 8 and 6, the green lines for $z$=4, 2, 1, 0.8 and 0.6, the
orange lines for $z$=0.4 and 0.2, finally the blue dashed line is for $z$=0.   }
\label{sfr_mass_redshift}
\end{figure}

It is also interesting to see the contribution to the   SFRD by galaxies
of different mass at different  redshifts. This is displayed in Fig.
\ref{sfr_mass_redshift}, where we show the product $SFR(z, M_{G}) \times N(z, M_{G})$ as
a function of the
total galaxy mass $M_G$ and  redshift. Each line is at constant redshifts. The color code
bins the lines in
three groups of redshift (namely, the red
lines are
for redshifts $z$=20, 15, 10, 8 and 6, the green lines for $z$=4, 2, 1, 0.8 and 0.6, the
orange lines for $z$=0.4 and 0.2, finally the blue dashed line is for $z$=0).  At high redshifts, the
dominant contribution is from the low-mass galaxies,
it shifts to that from higher mass galaxies at intermediate redshifts, and gradually it goes back
to the low-mass range for redshifts tending to zero.
Looking at the case $z$=0 in the $[SFR(M_G)\times N(M_G)]$ vs $M_G$ plane (Fig. \ref{sfr_mass_redshift}),
the slope
  $dlog[SFR(M_G) \times N(M_G)] / dlog M_G \simeq -0.3$
for $M_G$ passing from $10^8$ to $10^{12}\, M_\odot$, i.e. it mildly decreases with the galaxy mass.

\begin{figure}
\begin{center}
\includegraphics[width=7.5cm]{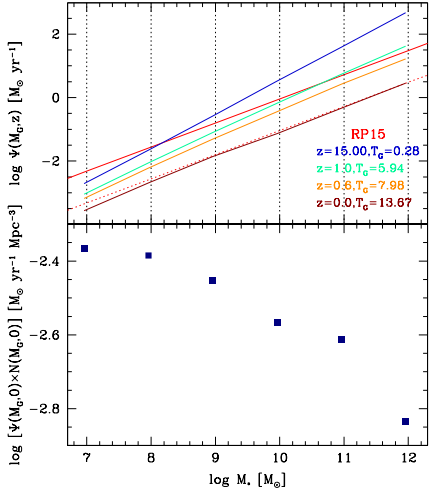}
\caption{ \textsf{Top Panel}: The SFR vs the galaxy stellar mass $M_s$
at different epochs in galaxies of different BM mass.
Since these models are without galactic winds,
$M_s \simeq 0.97\times M_{BM}$. The blue solid line is for $z$=15, the green line for $z$=1.0,  orange line
for $z$=0.6, and finally the dark red line for  $z=0$.
The solid red
line labeled RP15 are the observational data from \citet{RenziniPeng2015} for active galaxies.
Finally 
the red dotted
line is the RP15 decreased by factor of 10 (see text for details).
\textsf{Bottom Panel}: the product $SFR(0)\times N(M_G,0)$ vs mass in stars $M_s$ of each galaxy
in comoving $Mpc^3$.  }
\label{sfr_mass_number}
\end{center}
\end{figure}

To strengthen the above conclusion we look at the   correlations of the SFR and SFRD(z) with the  star mass
$M_s$ and/or $M_{BM}$ as a function of the redshift. Thanks to the high efficiency of star formation ($\nu$)
in all the models, $M_s \simeq 0.97 M_{BM}$.   The relationships in question are shown in the two panel of
 Fig. \ref{sfr_mass_number}.
The top panel shows the  SFR vs  $M_s$ at different redshifts,
whereas  the bottom panel displays the SFRD(z) at $z$=0.
The slope and zero point of the SFR vs $M_s$ relationships change with the redshift. The slope gets
smaller at decreasing
redshift: specifically  at $z$=15 indicated by the blue line,
$z$=1.0 and $z$=0.6 (the  green and orange lines respectively) and the $z$=0 (the dark red line).
These theoretical relationships are compared by the observational one by
\citet[][]{RenziniPeng2015}, the
 red solid line labelled RP15. The theoretical relation at $z$=0 
has the same slope of the one by \citet{RenziniPeng2015} and differs in the zero point. It coincides
with the RP15
lowered by a factor of 10 (red dotted line).  Model galaxies
at $z$=0 have  a minimal value of star formation, therefore they belong either to the group
so-called ``green valley'' galaxies or even to that of   quiescent objects. They could rise 
to the values of
\citet{RenziniPeng2015} by allowing the formation redshifts to span a wider range of values.  The key
result of this panel is that the slope of the SFR vs $M_s$ relation is the same as that of the
observational data over a wide range of redshifts ($0 \leq z < 1$).
In the bottom panel the product  SFR$\times N(M_G,0)$ decreases at increasing star mass of the galaxies.
Furthermore, there may be a 
qualitative agreement with the data of Fig.8 of \citet{Speagle_etal2014} (their Figure 8), who 
find that the  Main Sequence slope of the star-forming galaxies  increases with the redshift, 
i.e. the conversion of gas in stars 
decreases with time for all masses, the massive ones in particular.
This feature, otherwise known as ``downsizing'' from the observational point of view,
appears after a mere application of the $N(M_G,z)$, thus
it perfectly agrees with concordance
$\Lambda$-CDM cosmogony.
Finally, there is a point to note in the bottom panel: at a first look,
 the case with $M_s=10^{11} \, M_\odot$
 seems to deviate from the expected trend due to its apparently higher value. To single out the 
cause of it is a cumbersome affair.
Most likely it is due to inaccuracy in the derivation of the galaxy number densities for galaxy masses
in the high mass hand of the distribution. Although it is not in plain contrast with other values 
of stellar mass,
we plan to highlight the issue investigating other halo mass
functions in literature \citep[see][and references]{Murray_etal2013}

We conclude this section with the provisional result that our simplified model for the evolution of the
SFRD(z) nicely agrees with the observational data. However, it would be of general interest to single 
out the
physical ingredients that are ultimately responsible of this result. This will be the subject of a few ad
hoc designed experiments that are shortly illustrated in the following sections.

\begin{figure}
 \centering{
 {\includegraphics[width=7.5cm]{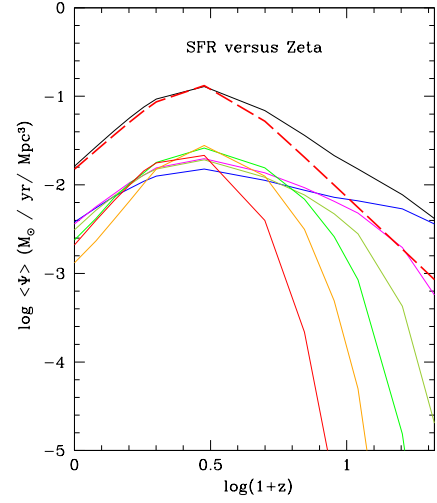} } }
 \caption{The contribution to the total SFRD(z) of Fig. \ref{sfr_madau} from galaxies of different mass.
 The top solid
line is the theoretical SFRD(z) whereas the dashed line is the analytical fit of the data by
\citet{MadauDickinson2014}. Finally the remaining lines are the partial contributions the the SFRD(z)
by galaxies of different BM mass from $10^{12} \, M_\odot$ (bottom) to $10^7\, M_\odot$ (top).     }
 \label{dissect_sfrd}
\end{figure}

\subsection{Dissecting the SFRD(z)}\label{dissect}

The first  test to
perform is to dissect the total $SFRD(z)$ in its components, i.e to single out the functions
$SFRD(z)_{M_{G}}$ whose sum at each $z$ yields back the total $SFRD(z)$. These are shown in
Fig.\ref{dissect_sfrd}. Remarkably,  all functions peak at
$z \simeq 2$. The decrease of the partial SFRDs at both sides of the peaks
cannot be attributed only to  number density (see
the curves in the \citet{Lukicetal2007} and  our Fig. \ref{fig_lukic})
because either they  are still
increasing toward their peak value (case of the high mass galaxies) or they have already reached their peak
at higher redshifts ($z \simeq 5$) as in the case of the low mass galaxies (this mirrors the combined 
effects
of the gravitational collapse and their destruction in the hierarchical aggregation). The only
 plausible explanation is that the SFRD(z) peak mirrors the superposition of the $SFR[(t(z)]$s
in existing galaxies
 which reach their peak in their star forming activity roughly at the same time. In more detail,
back in the past ($z > 5-6$) the dominant contribution came from low-mass objects; around the peak
interval all galaxies contribute in
nearly equal amounts even though those with masses in the range $10^{10}$ to $10^{11}$ $M_\odot$ are
more important; finally the low-mass ones are again dominating the contribution at low redshifts
($z < 1-2$).

\begin{figure}
\centering{
{\includegraphics[width=7.5cm]{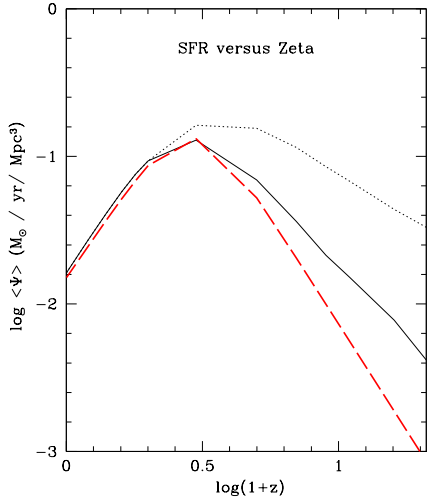} } }
\caption{Changing the relation $M_{BM} = \beta^{-1} M_{DM}$. The black dotted line is
the SFRD(z) derived from the arbitrary
assumption that $\beta=1$, equal amounts of DM and BM per galaxy. The red dashed line is the
 analytical best fit of the observational data of \citet{MadauDickinson2014} and  the solid black line
the theoretical SFRD(z) of Fig. \ref{sfr_madau}, obviously obtained with $\beta=6$.  }
 \label{BM_DM_re}
\end{figure}

\subsection{Changing the ratio $M_{DM}$ to $M_{BM}$}
It is worth examining the effect of adopting a different ratio $\beta = M_{DM}/M_{BM}$.
Among the various possibilities, there is one particularly interesting, i.e.
$\beta= 1$: DM and BM are in equal amounts.
The effect of this assumption on the first and second building blocks (the number of DM halos
of given mass as function of the redshift, and the galaxy chemical models) are easy to foresee.
The $N(M_{DM}, z)$ distribution remains the same and no particular remark has to be made.
The chemical models describing the evolution of the BM component within the DM halos remain unchanged
at least with our simplified picture in which the dynamical interaction of DM and BM is neglected.
Some effect would occur on the onset of galactic winds (if present) because  the gravitational potential
of the gas depends on the ratios $M_{BM}/M_{DM}$ and $R_{BM}/R_{DM}$;  the effect is however small, not
exceeding a factor of a few percent.
The major difference caused by the new  the relationship between BM and DM shows up when calculating
$[SFRD(z)]_T  = \sum_i N(M_{DM}, z) SFR_{i}(M_{DM}, z) {\Delta z}$ because the
$SFR_{i}(M_{DM}, z)$ that was the SFR of the BM galaxy with mass $M_{BM}=M_{DM}/\beta$ now is the
one of the BM galaxy with $M_{BM} = M_{DM}$: there can be a large factor in between that depends
on the redshift.
 In other words, at given total mass $M_G$ there is more BM to consider, so the SFR is
 more intense at early epochs.
The new cosmic SFRD(z) and the comparison of it with the reference one and the observational SFRD(z) of
\citet{MadauDickinson2014}  is shown in Fig. \ref{BM_DM_re}. The new SFRD(z) much resembles  one
of the reference case, nearly coincides with it on the tail from z=1 to z=0, but after that it flattens out
and runs well above the reference case (it peaks at about $z$=3
instead of $z$=1 to $z$=2 and runs above it by a factor of about three at higher redshifts).
One is tempted to argue that the cosmic SFRD(z) could be a good tracer of the amount of DM with respect
to BM.

\begin{table*}
\begin{center}
\caption{Characteristic quantities of the models  of Group B at the onset of the galactic wind.
The following quantities are shown: the baryonic mass $M_{BM}$ in
solar units, the accretion timescale $\tau$  in Gyr, the efficiency of star formation $\nu$,
the time $t_{GW}$ in Gyr
of the occurrence of the stellar wind, the gas fraction  $G_{g,GW}$, star mass fraction  $G_{s,GW}$,
the current metallicity $Z_{GW}$, the mean metallicity $<Z_{GW}>$, the SFR $SFR_{GW}$,
the gas
gravitational potential energy $\Omega_{g,GW}$, and the gas thermal energy $E_{th,GW}$ at the onset of
the galactic wind (both are per unit mass of the galaxy and in $erg\, g^{-1}$). The top models refer to
the case of standard rate of star formation
and condition for galactic wind. The bottom models refer to case in which the thermal budget given to the
interstellar medium is artificially cooled down to suitable value and at the same time the SFR is lowered
by means of $\nu_{eff}$  so that the galactic wind can occur only at the present time. See the text for
details.  }
\label{keydata_winds}
\begin{tabular}{ |r  c| c| c|  c| c| c|  c| c| c|  c|}
\hline
\multicolumn{11}{c}{Standard Galactic Winds and SFR}\\
\hline
\multicolumn{1}{|r}{$M_{BM}$  }      &
\multicolumn{1}{c|}{$\tau$    }      &
\multicolumn{1}{c|}{$\nu$     }      &
\multicolumn{1}{c|}{$t_{GW}$  }      &
\multicolumn{1}{c|}{$G_{g,GW}$}      &
\multicolumn{1}{c|}{$G_{s,GW}$}      &
\multicolumn{1}{c|}{$Z_{GW}$  }      &
\multicolumn{1}{c|}{$<Z_{GW}>$}      &
\multicolumn{1}{c|}{$SFR_{GW}$}      &
\multicolumn{1}{c|}{|$\Omega_{g,GW}$|} &
\multicolumn{1}{c|}{$E_{th, GW}$}    \\
\hline
1.0$\times 10^{7}$  &  6 & 10 & 0.007 & 0.010 & 0.001 & 0.0001 & 0.0001 &1.05E-03 & 1.72E-02 & 9.13E+00 \\
1.0$\times 10^{8}$  &  5 & 10 & 0.007 & 0.011 & 0.001 & 0.0001 & 0.0001 &1.06E-02 & 1.75E+00 & 9.27E+01 \\
1.0$\times 10^{9}$  &  4 & 10 & 0.010 & 0.011 & 0.001 & 0.0001 & 0.0001 &1.09E-01 & 1.80E+02 & 9.15E+02 \\
1.0$\times 10^{10}$ &  3 & 10 & 0.010 & 0.012 & 0.001 & 0.0008 & 0.0008 &1.21E+00 & 2.01E+04 & 2.13E+04 \\
1.0$\times 10^{11}$ &  2 & 10 & 0.100 & 0.030 & 0.019 & 0.0135 & 0.0083 &2.95E+01 & 5.22E+06 & 5.24E+06 \\
1.0$\times 10^{12}$ &  2 & 10 & 1.010 & 0.040 & 0.362 & 0.0385 & 0.0311 &4.03E+02 & 5.17E+08 & 5.32E+08 \\
\hline
\end{tabular}
\begin{tabular}{ |r  c| c| c|  c| c| c|  c| c| c|  c|}
\multicolumn{11}{c}{Modified SFR and  Conditions for the onset of  Galactic Winds }\\
\hline
\multicolumn{1}{|r}{$M_{BM}$  }      &
\multicolumn{1}{c|}{$\tau$    }      &
\multicolumn{1}{c|}{$\nu$     }      &
\multicolumn{1}{c|}{$\eta_{th}$  }   &
\multicolumn{1}{c|}{$G_{g,GW}$}      &
\multicolumn{1}{c|}{$G_{s,GW}$}      &
\multicolumn{1}{c|}{$Z_{GW}$  }      &
\multicolumn{1}{c|}{$<Z_{GW}>$}      &
\multicolumn{1}{c|}{$SFR_{GW}$}      &
\multicolumn{1}{c|}{$\Omega_{g,GW}$} &
\multicolumn{1}{c|}{$E_{th, GW}$}    \\
\hline
1.0$\times 10^{7}$  &  6 & 10 & 0.00001 & 0.033 & 0.920 & 0.0440 & 0.0357 &2.89E-04 & 1.87E-01 & 1.58E-01 \\
1.0$\times 10^{8}$  &  5 & 10 & 0.00010 & 0.020 & 0.928 & 0.0445 & 0.0365 &2.14E-03 & 1.12E+01 & 9.07E+00 \\
1.0$\times 10^{9}$  &  4 & 10 & 0.00500 & 0.040 & 0.903 & 0.0488 & 0.0348 &1.58E-02 & 2.24E+03 & 2.07E+03 \\
1.0$\times 10^{10}$ &  3 & 10 & 0.01000 & 0.007 & 0.928 & 0.0485 & 0.0371 &6.14E-02 & 3.99E+04 & 3.39E+04 \\
1.0$\times 10^{11}$ &  2 & 10 & 0.10000 & 0.007 & 0.923 & 0.0514 & 0.0370 &3.42E-01 & 2.79E+06 & 2.54E+06 \\
1.0$\times 10^{12}$ &  2 & 10 & 0.30000 & 0.008 & 0.913 & 0.0556 & 0.0366 &1.51E+00 & 9.24E+07 & 8.45E+07 \\
\hline
\end{tabular}
\end{center}
\end{table*}

\begin{figure}
\centering{
{\includegraphics[width=7.5cm]{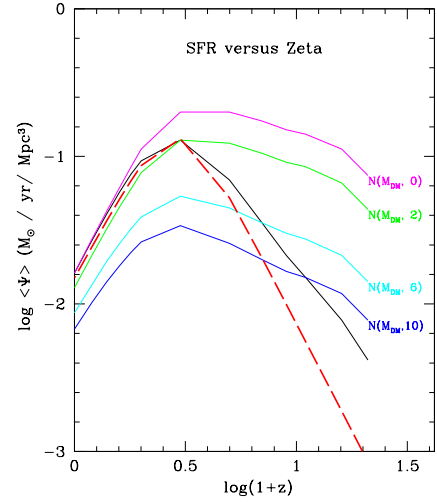} }  }
\caption{Changing the  $N(M_{DM}, z)$ relationship. See the text for details.   }
 \label{NDM_zeta}
\end{figure}

\subsection{Changing the $N(M_{DM}, z)$ relationship}

At each redshift the  gravitational aggregation of lumps of DM and BM in objects of larger
and larger total mass is described by the function
$N(M_{DM+BM}, z)$ whose mass dependence is customarily approximated to $N(M_{DM},z)$ thanks to
the large ratio $M_{DM}/M_{BM}$. However the exact shape of  the function $N(M_{DM},z)$ is still
uncertain, even if the one we have adopted may be a good approximation of the real one.
Basing on these considerations,
 it comes naturally to pose the question: how would the cosmic SFRD(z)
change if the underlying mass function of DM halos  was different from the one currently in use?
To answer the question without venturing in
arbitrary speculations, we perform a simple numerical experiment. We  assume that the mass distribution
does not vary with the redshift but only with mass, and test four mass distributions:
namely
$N(M_{DM},0)$, $N(M_{DM},2)$, $N(M_{DM},6)$ and $N(M_{DM},10)$.  Therefore, while the rate of star formation
in the model galaxies varies with the redshift,  their number does not. With this recipe,  we
 calculate the corresponding SFRD(z). We will refer to this as the \textit{false SFRDs}.
The results are
shown in Fig. \ref{NDM_zeta}. No renormalization has been applied to the false SFRD(z) to make
their peak value to coincide with  the observational one by
\citet{MadauDickinson2014}. It is interesting to note that  the false SFRDs  resemble
the real one at low redshift ($z \leq 2$),  strongly
deviate from it at intermediate  redshifts, and eventually tend again to it
at high redshifts.   Since the SFRs of the model galaxies are the same as those
of the reference frame, this clearly shows that $N(M_{DM}, z)$ is the  term that mainly  drives  the
shape of the cosmic SFRD(z).  The gravitational building up of galaxies at early epochs ($z \geq 1-2$) yields
the rising branch whereas in more recent epochs
($z \leq 1-2$) the  declining of the  mean SFR in galaxies  by gas consumption most likely prevails.

\begin{figure}
\centering{
{\includegraphics[width=7.5cm]{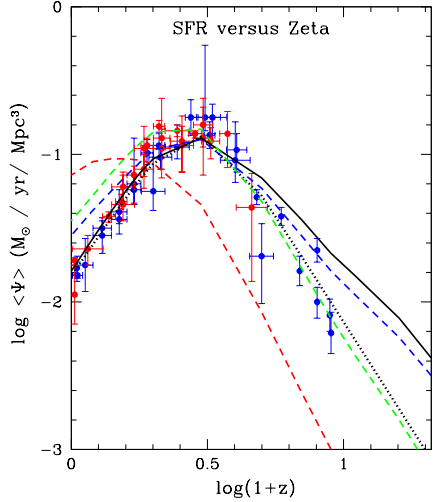} }  }
\caption{The effect of the intrinsic efficiency $\nu$ on the SFRD(z) derived from models of the
groups C  (dashed lines)  and B (solid line) and  the
comparison with observational data (filled circles with error bars) and their analytical fit (dotted line)
of \citet{MadauDickinson2014}. The three lines for the models of group C refer to  the different $\nu$
under consideration (0.1, 1 and 10 from the bottom to the top of the figure).}
\label{sfr_nu}
\end{figure}

\subsection{Changing the efficiency of star formation $\nu$}
The rate of star formation we have adopted contains also the efficiency parameter $\nu$ whose effects
are worth being investigated.
Figure \ref{sfr_nu} shows the SFRD(z) expected for models of type C in which the
efficiency parameter $\nu$ of the SFR is decreased from $\nu=10$ (top long dashed line)
to $\nu=1$ (middle long dashed line) and even to
 $\nu=0.1$ (bottom long dashed line). Together with the observational data (filled circles) we plot
 the analytical
fit (dotted line) by \citet{MadauDickinson2014}, and finally the theoretical SFRD(z) for models B (the solid black line).
It is soon evident
that models with a SFR too low ($\nu=0.1$) can be ruled out because too far off compared to the
observational data. The agreement between theory
and observational data is good for case B ($\nu=10$) and also case C models with high efficiency of SFR
$\nu=1$  and $\nu=10$, both cases, however, being somewhat higher than observed on the descending
branch towards $z$=0.

\subsection{Changing the SFH of galaxies}

It has repeatedly been said that an important requisite to get the observed SFRD(z)
is that the star formation in galaxies starts very small, increases to a peak value and then declines because
of gas consumption. How legitimate is the  kind of temporal dependence of the SFR
we have been using so far? In other words can we obtain the same SFRD(z) using  different types of SFR?

To test this point, we explore here two different alternatives: (i)  in each galaxy the rate of star
formation is constant
and equal to a suitable value so that the desired amount of stars is obtained; (ii) the rate of
star formation is a mere exponentially decreasing function from a maximum value at the beginning to the
present-day value.

\begin{figure}
\centering
\includegraphics[width=7.5cm]{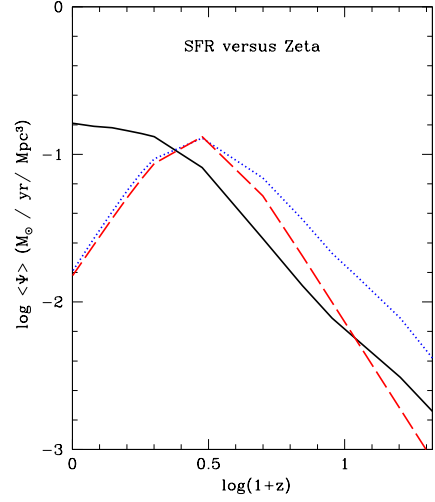}
\caption{The cosmic SFRD(z) predicted by galaxy models whose rate of star formation is constant
with time and equal to the mean value expected for models of group B (black solid line). The mean SFR is different
for each galaxy. The value is calculated as $M_s(T_G) = T_G^{-1} \int \psi(t) dt = <\Psi(t)> T_G$,
 where  $M_s(T_G)$
is the present day mass in stars of the galaxy with the same total mass but normal-time-varying
$\Psi(t)$, $T_G$ the present day age of the galaxy.  The blue dotted line is the SFRD(z) for models of group B
and the red dashed line is the observational fit by \citet{MadauDickinson2014}. }
\label{sfr_const}
\end{figure}

\textsf{Constant Star formation}. The analysis is made by means
of models B for which we calculate the mean SFR as
\begin{equation}
<SFR> = \frac{ \int_0^{T_{G}} \Psi(t) dt} {T_{G} }  \Rightarrow    M_s(T_G) = <SFR> \times T_{G}
\label{mean_sfr}
\end{equation}
where $T_G$ is the galaxy age,  $\Psi(t)$ the current SFR, and $M_s(T_G)$ the total mass in
stars
at the galaxy age $T_G$. All these quantities are known from the previous calculation of models B.
Since no galactic
winds are considered in models of group B and nearly all BM mass is converted into stars ($\simeq 90 \%$),
for all practical purposes their $<SFR>$ can be estimated inserting in eq.(\ref{mean_sfr})
$M_s(T_G) = 0.9\times M_{BM}$, $T_G \simeq 13.5$ Gyr, and expressing it in $M_\odot\, yr^{-1}$.
With the
aid of these $<SFR>$, we derive the new SFRD(z) with the usual procedure
and compare it with the observational one. The result is shown in   Fig.\ref{sfr_const}.
As expected, now the cosmic SFRD(z) simply increases with decreasing redshift, thus mirroring the
underlying increasing mean number density of galaxies of different mass. This finding lends strong support
to our previous conclusion about the  time dependence of the SFR taking
place in  each galaxy.

\begin{figure}
\centering{
{\includegraphics[width=7.5cm]{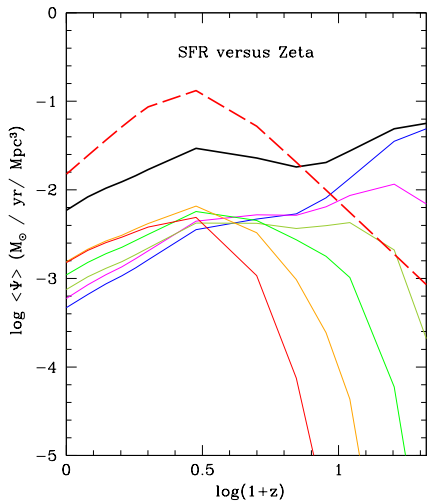} }    }
\caption{The predicted SFRD(z) for the closed-like models of type A (see text for details). The thin lines show
the partial contribution to the SFRD(z) from galaxies of different mass: from top to bottom the galaxies
are labelled by their BM mass scale from $10^7$ to $10^{12} \, M_\odot$.  The heavy solid line is the total
SFRD(z).
Finally, the dashed line is the  analytical fit of the data by \citet{MadauDickinson2014}  }
 \label{quench}
\end{figure}

\textsf{Exponentially decreasing star formation rate}.
To cast light on this issue we make use of the models of group A. In all these models the timescale of
mass accretion and intrinsic star formation efficiency are  $\tau= 0.01$ Gyr and $\nu=10$. These models
closely mimic the closed-box approximation. With these assumptions, the SSFR of these models
is essentially a simple exponential law. Consequently, the maximum star formation occurs at the beginning of
the SFH and declines ever since. The resulting SFRD(z) is shown in Fig. \ref{quench},
which shows the partial contribution to the SFRD(z) from galaxies of different mass (thin lines),  the
total SFRD(z), heavy solid line,  and  the observational fit.
Looking at the total SFRD(z) we note that it is very high (actually much higher than the observational
one) at high redshift, it has a lull at intermediate values and it remains lower than the observational
one at low redshifts. The reason for this awkward behavior can be accounted for by examining the
partial contribution from galaxies with different mass. First of all,  galaxies with
$M_G > 10^{10}\, M_\odot$ at decreasing redshift first increase, reach a peak value and then decrease again.
At z=0 their contribution is comparable within a factor of five. The galaxies of lower mass have, at
low redshift, contribution smaller than the massive ones, the opposite happens at high redshift, and
they are also
responsible for the intermediate redshift lull.  Finally, this can be attributed to the time dependence
of the SSFR in each galaxy which is simply a mere exponential law, the same for galaxy models.

The net conclusion of this experiment is that an always decreasing SFR from an
initial maximum to the present day value cannot generate the desired SFRD(z) unless other physical effects
are introduced.

 \begin{figure}
\centering{
{\includegraphics[width=7.5cm]{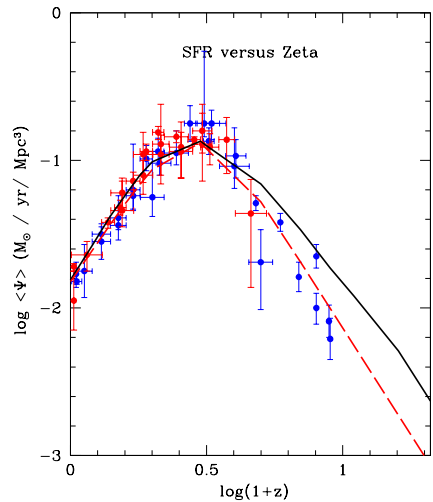} }   }
\caption{The predicted SFRD(z) for models of type B with galactic winds whose
key data are
reported in Table \ref{keydata_winds} and comparison with the observational data (filled circles) and
their analytical fit (dashed line) by \citet{MadauDickinson2014}.  }
 \label{wind}
\end{figure}

\subsection{Introducing galactic winds}
All the galaxy models used so far have been calculated ignoring the possible
presence of galactic winds (i.e. the condition (\ref{eth_omg})
for the onset of galactic winds has not been applied). In  this section, we take the  energy
injection by supernova explosions and stellar winds into account and apply condition (\ref{eth_omg}).
In this view of the whole issue of galactic winds, the above prescription implies that  when
condition (\ref{eth_omg}) is
verified, the remaining gas is supposed to escape the galaxy and further star formation does no longer
occur. The evolution of the remnant galaxy is a passive one and all the gas shed by  stars formed in the
previous epochs either in form of stellar wind or supernova explosions does no longer generate new stars.
In addition to this, owing to the different gravitational potential well of massive galaxies with respect
to the low-mass ones, the time at which the threshold energy for galactic winds is reached occurs earlier
in low-mass galaxies than in the massive ones. All this is inherent to the \citet{Larson1974} model of
galactic winds
which has been superseded by more sophisticated treatment of the wind process with the aid of NB-TSPH models
of galaxy formation and evolution \citep{ChiosiCarraro2002, MerlinChiosi2006,MerlinChiosi2007,Merlinetal2012}.
To illustrate the point, we show in the top part of Table \ref{keydata_winds} a few key quantities for models
 of Group B evolved in
presence of galactic winds according to the straight prescription of \citet{Larson1974}. With this 
prescription,
galactic winds occur very early so that the stellar content of a galaxy is hardly made. The problem 
can be partly cured either
by decreasing the efficiency of star formation (lower values of $\nu$) or by invoking a lower quantity
of energy actually injected by supernova explosions and stellar winds
to the interstellar medium (more efficient cooling of this energy).  Since in doing this a certain degree of
arbitrariness is unavoidable owing to the lack  of suitable constraints on the galaxy models in use,
we prefer to adopt a different strategy.

This modeling of the galactic winds is not realistic because the numerical NB-TSPH simulations
have indicated that galactic winds are not instantaneous
but take place on  long timescales. Gas heated up by supernova explosions and stellar winds
 and cooled down by radiative processes  not only gradually reaches the escape velocity but
also affects the efficiency of star formation because   the hot gas  is
continuously subtracted.
All this cannot be easily incorporated in the simple-minded galaxy model we are
using here. To cope with this difficulty we modify our model as follows.

First of all, feeling that the cooling algorithm we are using is not as good as the one currently
adopted in NB-TSPH models, we introduce an efficiency parameter $\eta_{th}$ ranging from 0
(no energy feed-back)  to 1 (full
energy feed-back) and accordingly change the  (\ref{eth_omg}) to the new one
\begin{equation}
\eta_{th} \times E_{th}(t) \geq |\Omega_{g}(t)|
\label{new_th_cond}
\end{equation}

Second we change the star formation law redefining the parameter $\nu$ as an effective efficiency given by

\begin{equation}
 \nu_{eff} = \nu \times \frac{|\eta_{th} \times E_{th} - |\Omega_{g}||}{\eta_{th} \times E_{th} + \Omega_{g}}
\label{nu_eff}
\end{equation}
where $\nu$ is the usual efficiency. By decreasing the efficiency of star formation at increasing $E_{th}$
we intend to mimic the fact that hot gas is likely less prone to generate stars by gravitational collapse.
 As consequence, the threshold stage for the onset of
galactic winds may occur much later in time or even avoided at all. Less gas is turned into stars as
if part of the gas is continuously escaping from the galaxy. The net SFR decreases with obvious consequences
on the SFRD(z).

New models of Group B are calculated using  the efficiency parameter
$\nu_{eff}$, and the new condition (\ref{nu_eff}) for the onset of galactic winds. The values of
$\nu_{eff}$ are chosen in such a way that the galactic winds occur
only at the present age or later. These parameters  are listed in the bottom part of Table
\ref{keydata_winds}. All over their history
these models have a  SFR lower than their standard counterparts thus mimicking the most important
effect of the energy feedback of evolving stars, i.e. heating up  part of gas and subtracting it to
star formation. The SFRD(z)
expected from these models and the comparison with the observational one of \citet{MadauDickinson2014}
is shown in  Fig.\ref{wind}. Theory and observations agree above all expectations.
Although our treatment of galactic wind is very crude, yet we suspect that galactic wind should only play a
marginal role on shaping the SFRD(z).

\begin{figure}
\centering{
{\includegraphics[width=7.5cm]{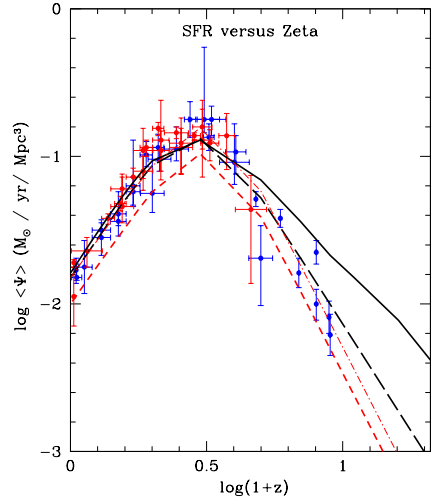} }   }
\caption{The theoretical SFRD(z) from models of group B (solid black line) compared with
the observational data (blue and red filled circles with error bars), the old analytical fit by  
\citet{MadauDickinson2014} (long dashed line),   the original empirical relationship 
by  \citet{MadauFragos2017} (short dashed line) and the same shifted by the factor 0.66 to compensate for 
the different assumptions for the stellar IMF (dashed dotted line). See the text for details.   }
\label{two_fits}
\end{figure}

\subsection{Changing the analytical best fit}
We conclude the analysis by comparing the theoretical models with new analytical best fit of the 
observational data by \citet{MadauFragos2017} who take into account recent data in the redshift 
interval ($4 \leq z \leq 10$0 and also the IMF by \citet{Kroupa2001} instead of the \citet{Salpeter1955}
one. Changing the IMF introduces a factor 0.66 passing from the old to the new one. 
The comparison  is shown in Fig.\ref{two_fits}. Correcting for this factor as appropriate, the 
agreement between theory and observation is still there.

\subsection{Comparison with N-Body cosmological simulations}

Recent attempts to model the  SFRD(z) in the framework of large scale simulations of hierarchical
galaxy formation in $\Lambda$-CDM cosmogony including both DM and BM have been made possible by the new
generation of numerical codes developed by \citet[][and references therein]{Hernquist_Springel_2003,
Springel_Hernquist_2003a,Springel_Hernquist_2003b,Vogelsberger_etal_2012,PuchweinSpringel2013,Baraietal2013},
in which much effort is paid to include radiative cooling and heating in presence of an UV background
radiation field, star formation and associated feedback processes. The  SFRD(z) in particular  has
been addressed by \citet[][see their Fig. 7]{KatsianisTescarietal2017} and \citet{Pillepich_etal_2017}.
The key results are the comoving mean SFR and cosmic SFRD as functions of look-back time and/or redshift
that are much similar to those we have  used here. It is worth emphasizing  that the mean SFH and
SFRD(z) refer to the whole slab of Universe under examination, and not to any galaxy in particular.

In our study we have taken a different perspective: starting from  galaxies of which we follow in
 detail the SFH, we integrate over the whole population of galaxies in the same
Universe slab (which
number is derived from the hierarchical growth of
structures in the $\Lambda$-CDM cosmogony) and we derive the total SFRD(z).



In other words, starting from individual objects, we reconstruct
the mean SFRD(z). In this context, the results of the present study are in perfect agreement with those
obtained from extensive and time consuming cosmological simulations.
The novelty of the present study is that we arrive at the same conclusions with a much simpler
approach, in which all physical foundations of the cosmic SFRD can be changed and separately analyzed
with almost no computational time needed.

\section{General remarks and conclusions}\label{Conclusions}

Prior to any consideration, we point out that (a) the HGF and galaxy SFR  are the starring
actors of the whole problem and (b) no  specific assumption is made
 to force the galaxy models in use  to reproduce the cosmic SFRD (the choice of the their leading 
parameter is suggested by other independent arguments). 
Basing on the present  analysis,  we  may conclude:

(i) The shape of the SFRD(z) is primarily driven by the cosmic mass distribution of galaxies, i.e. the
function $N(M_{G}, z)$  in place at each
value of the redshift. The galaxy mass distribution function   in turn partly results from the growth
of primordial fluctuations to the
collapse stage,  and partly from the aggregation of existing objects with active or quiescent star formation 
into new ones of larger mass (the classical hierarchical view).

(ii) The second important ingredient is the rate of star formation taking place in individual galaxies.
Only the so-called time-delayed SFR, i.e. a rate of star formation that starts small,
grows to a maximum and then declines, can yield the desired SFRD(z). In the formalism of the
infall models, in which the BM component (in form of gas) flows into the gravitational
potential well of DM at
a suitable rate proportional to an exponential time dependence $\dot{M}_{BM} \propto exp(-t/\tau)$ and
 is gradually converted into stars by the law $\dot{M}_{s} = \nu M_g$, giving rise to the time-delayed star
formation $\dot{M_s} \propto  \frac{t}{\tau} exp(-t/\tau)$. This kind of SFR is able to
reproduce the one inferred in galaxies of different morphological type
\citep[see][]{Sandage1986,Thomasetal2005} and also the SFR resulting from detailed numerical NB-TSPH
simulations of galaxies \citep[][]{ChiosiCarraro2002,Merlinetal2012}.
Constant and exponentially declining  SFRs cannot yield the observed SFRD(z).

However, also the intrinsic efficiency of star formation (the parameter $\nu$) has an important role because,
together with the timescale of mass accretion, eventually  drives the temporal dependence of star formation
passing from the one peaking at early epochs (high values of $\nu$) to that more skewed towards the present
(low values of $\nu$) passing through the interesting case of nearly constant star formation. We plan
to better investigate this issue in a forthcoming study.

(iii) The best galaxy models to use are those of type B or even type C with minor adjustments with respect to
those in use here that tend to produce too high metallicites. The problem can be easily solved either by
simply changing the net metal enrichment per stellar generation (the parameter $\zeta$ in
eqn.(\ref{zeta})) to lower values or playing with the other model parameters $\tau$, $\nu$, and $\nu_{eff}$.
Since this issue is marginal to our discussion we leave it to future investigations.
However, the agreement show by type B and C models  imposes a strong constraint on the type of
star formation taking place in galaxies.
It cannot be too much diluted over the Hubble time but instead it should be peaked at early epochs.

(iv) At early and late epochs (i.e. high and low redshifts)  the major contribution to the SFRD(z) comes
from galaxies of relatively low mass, whereas at intermediate redshifts the contribution from
intermediate mass galaxies may parallel or even exceed that from the low-mass ones. Although always
present at all epochs,
 the contribution from high mass galaxies is always smaller than that from low and intermediate mass ones.

(v) The energy feedback to the interstellar gas is only due to supernovae and stellar winds, no AGN
has been considered.  Radiative cooling of the injected energy is taken into account albeit in a
simplified fashion. This point needs to be improved.
 The present galaxy models are not the best ones to investigate the effect of galactic winds because,
owing to the one-zone approximation the onset of galactic winds at a certain  time  means sudden
interruption of the star formation process, whereas in real galaxies and also in numerical 3D-simulations
of galaxy formation
and evolution, galactic winds take place locally and over very long timescales without halting star formation
in the whole system. To cope with this, we preferred to decrease the efficiency of star formation as the
thermal content
of the gas, despite the radiative cooling, tends to approach and eventually overwhelms  the
gravitational potential  energy of the gas. In general galactic winds, even if they improve the
overall agreement of the models with observational data, are found to play a secondary role
in the context of the temporal evolution of the cosmic SFRD(z).

(vi) The SFRD(z) does not represent the instantaneous SFR in individual galaxies
$\Psi_G[t(z)]$, but it measures the mean
SFR of the population of galaxies in a unit volume. Therefore, it mirrors  the product
$\Psi_{G}[t(z)]\times N[M_G (t(z)]$, where $M_G$ is the total mass of a galaxy and $t(z)$ is the particular
time-redshift relation of the cosmological
model of the Universe that is adopted.  Using the SFRD(z) instead of $\Psi_G[t(z)]$  to model the history
 of single
galaxies may lead to wrong results. The opposite is also true.

(vii) We have adopted the HGF of \citet{Lukicetal2007},  which in turn stems from the
HMF of \citet{Warren_etal_2006}, simply because it is an easy-to-use tool for our purposes. 
However, owing to the well known problem of the non-universality of the  fitting function $f(\sigma)$  
\citep{Tinker_etal2008}, other models for the HMF can be found \citep{Murray_etal2013}.  
We plan to investigate this issue by using different HMFs.

(viii) The present approach yields results that fully  agree with those from the highly sophisticated large 
scale  numerical simulations. Therefore it  should be considered as a complementary tool for exploring 
different assumptions  concerning basic physical processes 
such as the
star formation law and the nature and efficiency of the energy feedback.

(ix) We plan to refine the present modeling of the SFRD(z) history by replacing the simple
galaxy models with a library of
3D N-body simulations of galaxy formation and evolution and also the number density evolution
of galaxies of different
mass, i.e. the functions $N(M_G,z)$ with the aid of ad hoc designed Monte-Carlo simulations.
Finally, we will follow  the photometric evolution of
the galaxies to investigate the relationship between the SSFR and stellar
mass content in galaxies of different mass, redshift and colors.
   
(x) As  final conclusions we would like to shortly answer a few important questions 
that could be raised such as for instance:
Why is the SFRD(z) small at high and low redshift? Is the quenching of SF at $z \leq 2$ associated with a 
decreasing gas supply at late epochs? Why is star formation inefficient at early times even in the absence of feedback?
Why is it possible to reproduce the data without AGN feedback?
What is the meaning of the particular combinations of parameters $\nu$ and $\tau$ required to
reproduce the data?  

The time (redshift) dependence of the SFR in the model galaxies is 
the result of the cross-effect of two physical processes: the gas accretion at a suitable rate 
onto the galaxy potential well and the gas consumption by star formation according to a Schmidt-like law. 
By controlling these two parameters the galaxy models can be tuned to match the 
gross features of real galaxies all along the Hubble sequence.  
The key feature of these galaxy models is that independently of 
the galaxy mass the SFR starts small, grows to a maximum and then declines as function of time. However 
the same SFR is strong and peaked at early epochs in massive objects (having the
early type galaxies as 
counterparts), mild and prolonged in the intermediate mass galaxies (observational counterparts the 
disk galaxies), and very mild and likely stretching (perhaps in recurrent bursts of activity not 
considered here) all over 
the Hubble time (observational counterparts the irregular galaxies). As already mentioned, 
this scheme is strongly supported by the body of observational data  of galaxies and the N-body
simulations of these. This tuning of the galaxy models in usage here has been  made over the years  
independently of the cosmic SFRD issue.
At low redshift, the ``quenching'' of star formation, is simply caused by the fact individual galaxies 
tend to run out of fuel (gas) in the star forming activity. At high redshift, a similar trend is 
recovered because galaxies are still in the gas accumulation phase and little gas has already 
reached the threshold density 
required for star formation to occur \citep[it is worth recalling here that 
stars form in very dense environments]{Krumholz_2015}. So at these very early epochs, the 
natural expectation is the  star formation activity is low but growing with time. This trend would 
mimic the effect of some quenching at early epochs.     

Our reference SFRD(z) of Fig. \ref{sfr_madau} obtained
with standard energy feedback from
supernovae and stellar winds, with  no AGNs and no galactic winds, is
already in rather good agreement with the
observational one from z=0 to z=2 (the reference case simply mirrors the picture 
outlined above for the natural behavior of SFR in galaxies) whereas  it tends to depart 
from it at increasing redshift. At redshift 
$z\simeq 10$ it is about a factor of 2 to 3 higher than expected.  
The presence of galactic winds slightly improves the
agreement in the latter region (see Fig. \ref{wind}) and perhaps some other effects like mild 
quenching by AGNs could  completely remove the discrepancy. 
Our provisional  conclusion is  that strong and exotic quenching of the star
formation in the interval $2\leq z \leq 8)$  
\citep[see for instance][and references]{Tescarietal2014, Renzini2016} is not strictly needed. 
The only case in which either strong quenching and/or dust obscuration or both are required is when an
an exponential  SFR is used.
However, the resulting SFRD(z) differs from the
observed one in many other details and has to be discarded. Therefore, quenching does not likely
play an important role in
shaping the observed SFRD(z) as compared to the combined effect of the
HGF $N(M_{DM},z)$
and of the $\Psi(t)$ modulated by the gradual accumulation of gas within the total gravitational potential
well  and conversion of it into stars. It goes without saying that AGNs ad galactic winds are not excluded 
from the above picture, but simply they are suspected to play a role less important than customarily claimed.

\section*{Acknowledgments}
We would like to thank the anonymous referee for his/her useful critical comments that helped us to amend and improve 
the first version of the paper. C. Chiosi, F. Brotto, R. De Michele, and V. Politino are deeply grateful 
to the Heraeus Foundation for the financial support to attend the Heraeus Summer School 2016 
``Origins of Stars and Planets'' (August 2016, Florence, Italy) where this study was 
presented for the first time. C.C. would like to thank the Physics \& Astronomy Department of the 
Padova University for the kind hospitality and computing support.

\bibliographystyle{mn2e}

\bibliography{COSMIC_SFR_BIBLIO}

\label{lastpage}

\end{document}